\documentclass[fleqn,usenatbib]{mnras}

\usepackage{newtxtext,newtxmath}
\usepackage{xcolor}
\usepackage[T1]{fontenc}

\DeclareRobustCommand{\VAN}[3]{#2}
\let\VANthebibliography\thebibliography
\def\thebibliography{\DeclareRobustCommand{\VAN}[3]{##3}\VANthebibliography}

\usepackage{graphicx}	
\usepackage{amsmath}	

\definecolor{amaranth}{rgb}{0.9, 0.17, 0.31}

\usepackage{soul}	
\usepackage{xcolor}

\title[High-energy particles in starburst winds]{Particle acceleration and multimessenger emission from starburst-driven galactic winds}

\author[E. Peretti et al.]{
Enrico Peretti$^{1,2,3}$\thanks{E-mail: peretti@nbi.ku.dk},
Giovanni Morlino$^{4}$, 
Pasquale Blasi$^{2,3}$ and
Pierre Cristofari$^{5,2,3}$
\\
\\
$^{1}$Niels Bohr International Academy, Niels Bohr Institute,University of Copenhagen, Blegdamsvej 17, DK-2100 Copenhagen, Denmark\\
$^{2}$Gran Sasso Science Institute, Viale F. Crispi 7, 67100 L’Aquila, Italy\\
$^{3}$INFN/Laboratori Nazionali del Gran Sasso, Via G. Acitelli 22, 67100 Assergi (AQ), Italy\\
$^{4}$INAF, Osservatorio Astrofisico di Arcetri, L.go E. Fermi 5, I-50125 Firenze, Italy\\
$^{5}$Observatoire de Paris, PSL Research University, LUTH, 5 Place J. Janssen, 92195 Meudon, France \\
}

\date{Accepted XXX. Received YYY; in original form ZZZ}

\pubyear{2021}

\begin{document}
\label{firstpage}
\pagerange{\pageref{firstpage}--\pageref{lastpage}}
\maketitle

\begin{abstract}
The enhanced star forming activity, typical of starburst galaxies, powers strong galactic winds expanding on kiloparsec (kpc) scales and characterized by bubble structures. 
Here we discuss the possibility that particle acceleration may take place at the termination shock of such winds. 
We calculate the spectrum of accelerated particles and their maximum energy, that turns out to range up to a few hundred petaelectronvolt (PeV) for typical values of the parameters. 
Cosmic rays accelerated at the termination shock are advected towards the edge of the bubble excavated by the wind and eventually escape into extragalactic space. 
We also calculate the flux of gamma rays and neutrinos produced by hadronic interactions in the bubble as well as the diffuse flux resulting from the superposition of the contribution of starburst galaxies on cosmological scales. 
Finally, we compute the diffuse flux of cosmic rays from starburst bubbles and compare it with existing data.  
\end{abstract}




\section{Introduction}

Starburst galaxies (SBGs) are unique astrophysical objects characterized by an intense star formation rate (SFR), and a correspondingly higher rate of supernova (SN) explosions. Since SNe and winds of young stars are believed to be acceleration sites of cosmic rays (CRs), SBGs are likely to be powerful cosmic--ray factories. The star forming activity is often located in sub--kpc sized regions, known as starburst nuclei (SBNi)~\citep{Kennicutt:1998zb}, with rather extreme conditions: high gas density ($n \gtrsim 10^2 \, \rm cm^{-3}$), intense infrared--optical luminosity ($U_{\rm RAD}\gtrsim 10^3 \, \rm eV \,cm^{-3} $) and strong magnetic fields ($B \gtrsim 10^2 \, \rm \mu G$) are inferred in SBNi \citep{Gao_Solomon_2004,Mannucci_etal_2003,ForsterSchreiber_2001,Thompson_etal_2006,Papadopoulos}.
The level of turbulence is also expected to be very high because of the repeated SN explosions and stellar winds. This turbulence is likely to slow down the spatial transport of charged high--energy (HE) particles, which therefore lose most of their energy inside SBNi. We refer to this mode of transport as {\it calorimetric}, and its implications have been discussed in detail by \citet{Yoast-Hull_M82_2013,Peretti-1,Krumholz2020}.
Multiwavelength observational campaigns from radio to hard X--rays \citep[see e.g.][]{Radio_Williams_Bower_2010, Carilli_2006,Wik_NGC253_2014}, and especially the spectra inferred from observations in the gamma--ray range, indicate that the transport of HE particles is strongly regulated by energy losses \citep[see e.g.][]{Ackermann_Fermi_2012,Peng:2016nsx,Abdalla:2018nlz,Ajello2020,Kornecki2020,Kornecki2021,Werhahn2021}.

A peculiar aspect of SBGs is represented by the amount of target material for nuclear interactions, potentially leading to copious production of neutrinos and gamma rays. The contribution of SBGs to the neutrino flux measured by the IceCube Observatory~\citep{First_Ice_nu,IceCube2020_LAST} has been extensively discussed by many authors \citep[][]{Loeb-Waxmann2006,Tamborra-Ando-Murase:2014,Bechtol-Ahlers:2015,Sudoh,Palladino-starburst:2018,Peretti-2,Ajello2020,Ambrosone2020,Ambrosone2}, together with the compatibility of the predictions with existing constraints imposed by gamma--ray observations \citep{Ackermann_Fermi_2012,Lisanti_2016}. The seriousness of these constraints stimulated the search for powerful hidden CR accelerators in environments highly opaque to gamma rays and yet transparent to neutrinos \citep{Capanema2021} like the inner core of Active Galactic Nuclei \citep[AGNi; see e.g.][]{Murase-hidden-CR-ACC,Murase:2020} or to reconsider the contribution from an extended region around the Galaxy \citep[see e.g.,][]{Taylor-2014,Blasi-Amato2019,Recchia-M31}.

Recent anisotropy measurements performed by the Pierre Auger Observatory \citep{Auger-anis} support the idea that SBGs might play an important role in the production of ultra--high--energy cosmic rays (UHECRs) \citep[see also][]{Anchordoqui-Romero1,Anchordoqui2018}. 
This piece of information adds to previous indications of the existence of a CR component with light mass and possibly of extragalactic origin, in the energy region $\lesssim$~EeV \citep[][]{Tunka2016,Kascade-GRANDE2017,IceTop2019}.

Starburst winds have indeed been suggested to accelerate particles above PeV energies \citep[][]{Dorfi2012,Bustard_Zweibel2017} 
and subsequently produce photons through non--thermal processes \citep[][]{Romero-Muller,Buckman2020,Muller-Romero}. These phenomena, together with the calorimetric transport of CRs and the intense photon backgrounds in SBNi, led to the careful investigation of the emission and absorption of gamma rays in the central regions of SBGs, and the correlated neutrino emission \citep{Peretti-1,Peretti-2}.
Despite the potential importance of these astrophysical objects for a variety of phenomena, the modeling of the processes of acceleration and interaction of CRs in SBGs remains rather poor and yet it is crucial if to assess their role as sources of high energy radiation and CRs in a reliable way. 

As stated above, particles are not only accelerated in the nuclei of SBGs, but also in the (kpc--sized) wind structures expanding from the SBN region to the circumgalactic medium (CGM).
While in our previous works on SBGs \citep[][]{Peretti-1,Peretti-2} we focused our attention on phenomena occurring inside the SBN, here we discuss the starburst winds as potential additional sites for particle acceleration and interactions. 

Starburst winds are inferred to be powered by the mechanical energy and heat produced by SNe and young stars possibly combined with some contribution due to the radiation pressure \citep[see e.g.][]{Zhang2018}.
The intense activity heats and pressurizes the ISM \citep[see][for detailed observation of M82]{Westmoquette1,Westmoquette2} creating a hot cavity and eventually inflating a powerful thermally--driven wind bubble \citep[see][]{Veilleux}. 
Starburst winds are characterized by high mass--loss rate ranging from a few $\rm M_{\odot} \, yr^{-1}$ for moderate starbursts up to $10^2 \rm M_{\odot} \, yr^{-1}$ in Ultra Luminous Infrared Galaxies (ULIRGs) \citep[see][for details]{Cicone-2014} or starburst coexisting with (or replaced by) active galactic nuclei (AGNi) \citep[see e.g.][]{Lamastra1,Wang-Loeb2017,Liu2018,Lamastra2}. 

Measurements of the wind speed are often based on detection of spectral lines associated to the warm and cold phases of the ISM embedded in the wind bubble and indicate velocities of the order of hundreds of $\rm km \, s^{-1}$.
On the other hand, theoretical models and X--ray observations show that the hot phase of the wind has a much higher velocity of the order of $10^3 \, \rm km \, s^{-1}$ \citep[see e.g.][]{Strickland-Heckmann2009}. 
These fast outflows easily break out of their galactic disks and expand into the surrounding galactic halos \citep[see][hereafter CC85]{ChevalierClegg85}. 

Wind bubbles are characterized by an innermost region of fast and cool wind powered by a central engine. The fast wind region extends up to the wind termination shock (also referred to as the wind shock), where the wind plasma is slowed down and heated up. 
A forward shock expands into the circumgalactic medium, typically with transonic velocity. Between the two shocks the contact discontinuity separates the shocked wind from the shocked swept--up halo medium \citep[see e.g.][]{Koo-McKee}.
The starburst activity can last for hundreds of millions of years (Myr) thus potentially producing an approximately steady injection of particles during this time~\citep[see][]{DiMatteo2008,McQuinn2009,Bustard_Zweibel2017}. 

Here we investigate the process of diffusive shock acceleration (DSA) of particles at the wind termination shock of starburst--driven winds, and estimate the associated production of gamma rays and neutrinos produced in the entire bubble excavated by the wind, and the flux of protons escaping such bubble.
We adopt the semi--analytic approach to CR transport at the termination shock, as developed by \citet{MBPC2021} (hereafter MBPC21) for the case of winds associated to star clusters. This theoretical approach allows us to establish a direct connection between the environmental conditions in the wind and the particle acceleration process, with special attention for the maximum energy of accelerated particles. Moreover the transport of the non--thermal particles in the entire wind bubble is described rigorously, taking into account diffusion, advection, adiabatic losses and gains, as well as catastrophic energy losses. This enables us to calculate the cumulative contribution of starburst winds to the diffuse gamma--ray and neutrino fluxes exploring the associated proton flux that we could observe at Earth as CRs above the \textit{knee}.

Our investigation shows that: 1) protons can be accelerated up to hundreds of PeV at the starburst wind termination shock; 2) gamma rays and neutrinos are produced as secondary products of $pp$ and $p\gamma$ interactions in these systems, possibly leading to detectable spectral features; 3) the contribution of starbursts to the diffuse neutrino flux can be dominant without exceeding the diffuse gamma--ray flux observed by Fermi--LAT; 4) accelerated particles escaping starburst systems can provide a sizeable contribution to the  light CR component observed above the knee.

The structure of the article is as follows: in \S~\ref{Sec2: Qualitative} we provide a description of the wind bubble. In \S~\ref{Sec3: Model} we describe the modelling of acceleration and transport in the system, and provide the main details of our semi--analytical approach to CR transport. 
In \S~\ref{Sec4: Results - isolated galaxy} we discuss the solution of the transport equation and the corresponding maximum energy as a function of the relevant parameters. We also show the associated gamma--ray and neutrino fluxes and the flux of CR protons escaping the bubble, for some benchmark cases. 
In \S~\ref{Sec5: Results - diffuse} we explore the multimessenger potential of the combined contribution of wind bubbles in the context of the diffuse fluxes observed at Earth. 
In \S~\ref{Sec6: Discussion + Conclusions} we summarize our results and draw our conclusions.

\section{Evolution and properties of the wind bubble}
\label{Sec2: Qualitative}

The typical lifetime of a starburst event is of order $\sim 200-300 \, \rm Myr$~\citep{DiMatteo2008}: at formation the structure is fueled by energy and mass released by young OB and Wolf--Rayet stars for about $6 \, \rm Myr$. After this initial stage, the first core collapse SN explosions are expected to take place. The energy and mass that they release dominates over the ones due to the young stars activity. In the minimal assumption of an \textit{instantaneous} starburst trigger, the activity would run out in about $40$ Myr when $8 \, M_{\odot}$ stars end their life~\citep[see also][]{Veilleux}. In practice, the actual duration of a starburst is determined by the star forming activity, which can last up to few hundred million years, as mentioned earlier. Such time scale is much longer than the typical duration of the processes of particle acceleration and transport in the bubble produced by the starburst activity, so that from this point of view, SBGs and their wind superbubbles can be considered as steady state systems for HE particles \citep[see also][for related discussions]{ZIRAKASHVILI20061923,Bustard_Zweibel2017}.

The engine of a starburst--driven galactic wind is the activity of SNe and massive stars which heat and pressurize the interstellar medium (ISM) excavating a hot bubble where temperature and pressure are $T \sim 10^8 \, \rm K$ and $P/k_B \sim 10^7 \, \rm K \, cm^{-3}$ (as also discussed in CC85). 
Once the starburst event has started, the bubble expands above and below the galactic disk due to the pressure unbalance between its interior and the unshocked host galaxy ISM and eventually reaches the scale height of the disk, breaking out into the galactic halo. 
Inside the disk, instead, the bubble remains confined by the ISM pressure~\citep[see][]{Tenorio_Tagle_1997,Cooper2007}. 
As shown in recent numerical simulations \citep[][]{Fielding2018,Schneider2020}, the clustered activity of SNe typical of SBNi is strong enough to drive and sustain a powerful galactic outflow. 
In this framework, CRs could also contribute as a supplementary ingredient powering an outflow in very active star forming galaxies as discussed by \citet{Hanasz2013}. 
However, their importance in contributing to the wind launching is highly uncertain due to the possible impact of the dense and turbulent environment on their transport \citep[see e.g.][]{Krumholz2020} and their severe energy losses in the core of SBGs \citep[see e.g.][]{Peretti-1,Kornecki2021,Werhahn2021}.
On the other hand, in the case of a less intense and spatially extended star formation, typical of the spiral arms of mild star forming galaxies, where energy losses are usually negligible, the additional contribution of cosmic rays \citep[see e.g.][]{Breitshwerdt91,Everett2008,Recchia2016,Pfrommer2017,Girichidis2021} and radiation pressure may be necessary to launch a galactic outflow \citep[see e.g.][]{Zhang2018}.

The dynamics of starburst winds \citep[][]{Strickland-Stevens-2000,Strickland_2002} is qualitatively similar to that of stellar winds and winds of star clusters \citep{Castor-1976,Weaver77,Koo-McKee,Koo-McKee2} when the galactic ISM is roughly homogeneous \citep[][]{Strickland_2002}. However, when the medium is inhomogeneous, as expected in realistic cases \citep[][]{Westmoquette1,Westmoquette2}, the hot gas follows the path of least resistance out of the disk, resulting into a non homogeneous outflow. 
Once in the halo, the hot gas expands freely and the geometry can be reasonably assumed to be spherical \citep[see][]{Cooper2007}. For our purposes, the assumption of a spherical geometry is well motivated by the fact that accelerated particles probe large distances, averaging out any spatial inhomogeneities.

Radiative losses can affect the wind dynamics and several theoretical and numerical works investigated the possible role of such losses, leading to a wide range of possible scenarios \citep[see][and references therein]{Bustard-2016,Zhang2018}.
If the starburst wind is approximately adiabatic \citep[as shown in numerical simulations, see e.g.][]{Fielding2017,Schneider2020}, its behavior is in good agreement with the analytic model developed in CC85, and adopted in this work. 

The first stage of the evolution of the wind bubble is characterized by a free expansion which ends when the mass of the swept--up ambient medium becomes comparable to the mass injected in the form of a wind ($t_{\rm free} \lesssim 1 \rm \, Myr$ for an average halo density $n_h\approx 10^{-3} \, \rm cm^{-3}$). 
The wind is supersonic, so that it is preceded by a forward shock, while a reverse shock is launched towards the interior, the so-called termination shock. During the free expansion phase, the two shocks move outwards but staying very close to each other. The shocked wind and the shocked ISM are separated by a contact discontinuity. 
When the accumulated mass eventually becomes larger than the mass added in the form of a wind, the outflow decelerates appreciably. If the CGM is assumed to be spatially homogeneous, the radius of the forward shock changes in time as $R_{\rm FS} \propto t^{3/5}$, while the termination shock follows the trend $R_{\rm sh} \propto t^{2/5}$~\citep[see][]{Weaver77,Koo-McKee}. The bubble eventually reaches a pressure confined state, typically after a few tens of Myrs. 
This late stage of the evolution is characterized by a pressure balance between the cool wind ram pressure and the pressure of the undisturbed halo medium $P_h$ (which, in turn, is in equilibrium with the pressure of the shocked wind). 
At this point, the wind shock is stalled while the contact discontinuity and the forward shock keep slowly expanding in the CGM.
As detailed in \citet[][]{Lochaas2018} \citep[see also][]{Strickland-Stevens-2000}, the dynamics of the wind bubble depends on the density profile of the CGM gas. 

\begin{figure}
	\includegraphics[width=\columnwidth]{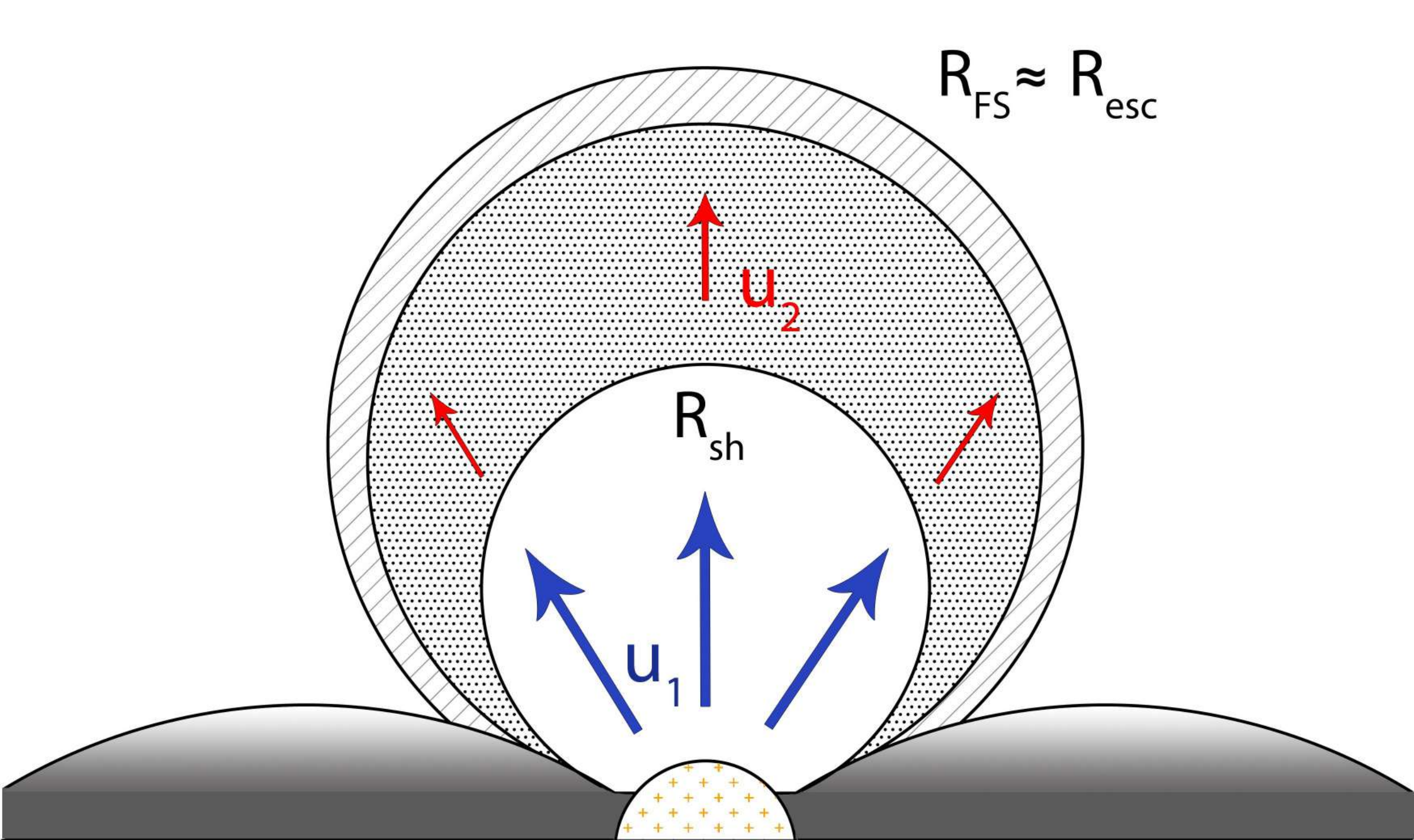} \quad \includegraphics[width=\columnwidth]{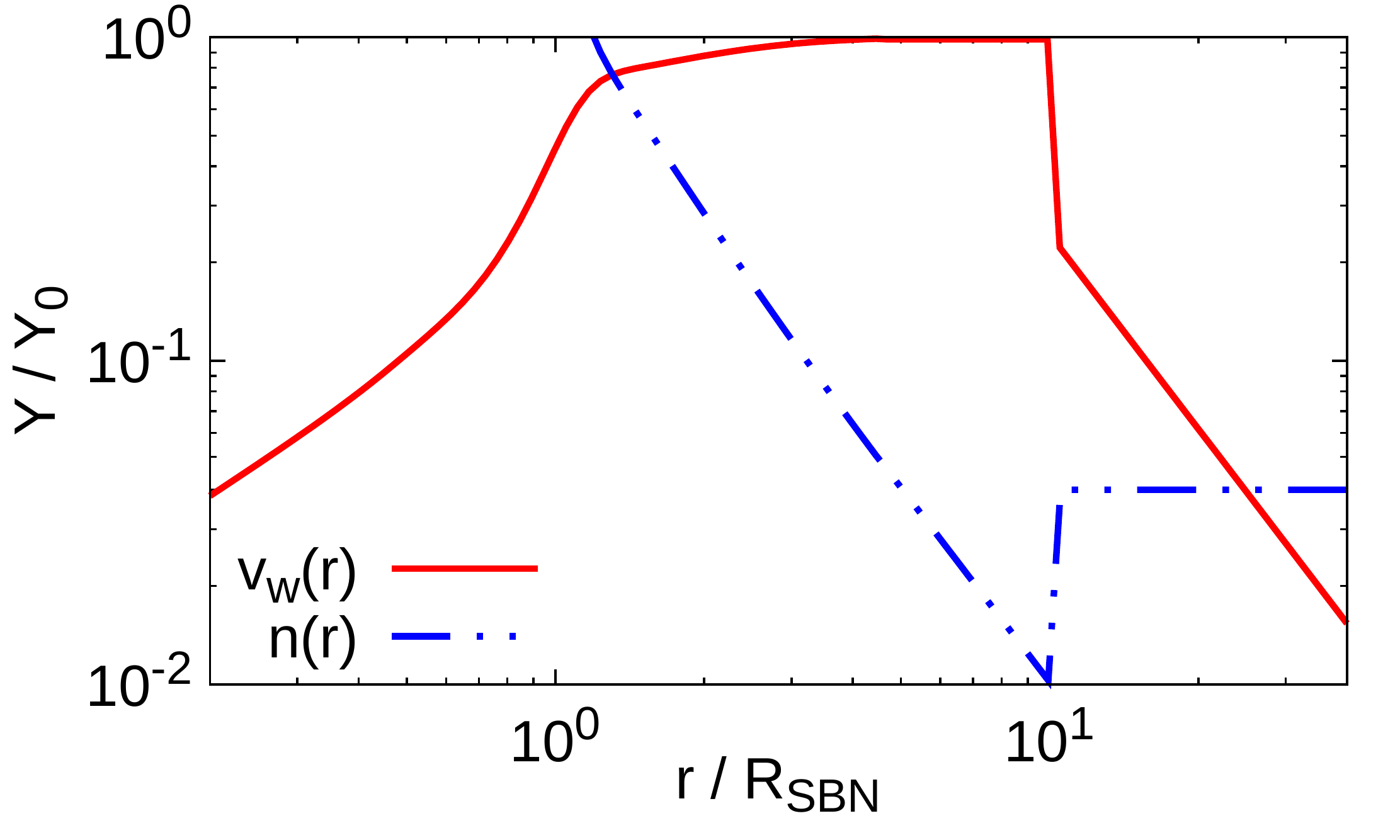}
    \caption{Top panel: structure of the wind bubble. The SBN, from which the wind is launched, is located in the center of the galactic disk. The blue (red) arrow corresponds to the cool (shocked) wind region. The wind shock ($R_{\rm sh}$) separates the two regions. The forward shock (at $R_{\rm FS}$) bounds the system from the undisturbed halo region (credit: I. Peretti). Bottom panel: wind profile (thick red) and particle density profile (dot-dot-dashed blue), where Y is the density or velocity, and $Y_0$ is the normalizing density or velocity. The plot is in arbitrary units for illustrative purposes. The location of the wind shock is assumed to be at $10 \, R_{\rm SBN}$ for illustrative purposes.}
    \label{fig:Image}
\end{figure}

The structure of the starburst--driven wind bubble can be pictured as onion--like (see top panel of Figure~\ref{fig:Image}). The SBN, responsible for launching and powering the outflow, is located at the center of the system. The wind speed increases approaching the boundary of the SBN, where it becomes supersonic and quickly reaches its terminal velocity ($V_{\infty}$). At this point the wind velocity remains basically constant (see CC85 and lower panel of Fig \ref{fig:Image}), up to the termination shock (located at $R_{\rm sh}$), where the wind is slowed down and heated up. 
As we discuss below, this configuration is very interesting from the point of view of particle acceleration, in that the upstream region is in the direction of the SBN, hence particle escape from the upstream region is inhibited and becomes possible only through the external boundary of the wind bubble. 
The medium in which a galactic wind bubble expands affects the spatial structure of the bubble. Galactic halos are inferred to be characterized by a hot diffuse gas component where typically $n_h \lesssim 10^{-2} \, \rm cm^{-3}$ and $T_h \sim 10^6-10^7 \, \rm K$ \citep[][]{Anderson2015-halos,Tumlinson2017}. Hence, in a starburst CGM the thermal pressure is expected to be $P_h/k_B \lesssim 10^5 \, \rm K \, cm^{-3}$ (where $k_B$ is the Boltzmann constant). 

In evolved wind bubbles, the balance between the thermal pressure in the halo and the wind ram pressure, $\rho \, v_w^2$, sets the position of the termination shock: 
\begin{equation}
    R_{\rm sh} \approx \sqrt{\frac{\Dot{M}V_{\infty}}{4 \pi \, P_h }} = 6.2 \, \Dot{M}_0^{1/2} \, V_{\infty,8}^{1/2} \, P_{h,4}^{-1/2}  \rm \, kpc.
    \label{Eq: shock-radius}
\end{equation}
where $\Dot{M}$ ($\Dot{M}_0$) is the wind mass loss rate (in units of $1 \, \rm M_{\odot} \, yr^{-1}$), $V_{\infty,8}$ is the terminal wind speed in units of $10^8 \, \rm cm \, s^{-1}$ and $P_{h,4}$ is the halo pressure in units of $10^4 \, k_B \, \rm cm^{-3} \, K $.
These three parameters characterize the global properties of the system~\citep[see also][for aditional details and connection to the core activity]{Veilleux,Strickland-Heckmann2009}.
While the termination shock is approximately stalled, the forward shock continues to expand as:
\begin {equation}
    R_{\rm FS} \approx 10 \, [\Dot{E}_{43} n_{\rm h,-3}^{-1}]^{1/5} \, t_7^{3/5} \, \rm kpc,
    \label{Eq: escape-radius}
\end{equation}
where $\Dot{E}_{43}= [\Dot{M}\,V_{\infty}^2/2]/10^{43} \, \rm erg \, s^{-1}$ is the wind power, $n_{\rm h,-3}$ is the halo density in units of $10^{-3} \, \rm cm^{-3}$ and $t_7$ is the time in units of 10 Myr \citep[see also][]{Koo-McKee}. It follows that the typical Mach number of the forward shock is of order unity, starting at times of order $\sim 10 \, \rm Myr$. This  is the main reason why efficient particle acceleration is not expected to take place at the forward shock.  

At the termination shock the conditions are more favorable. The temperature of the plasma at the wind shock sets the local sound speed.
The adiabatic expansion cools the gas as $T \propto r^{-4/3}$, so that assuming a SBN size $R_{\rm SBN} \sim 200$ pc and $R_{\rm sh}$ as given by Equation~\eqref{Eq: shock-radius}, one can expect a temperature $T \approx 10^6$ K at the wind shock when the SBN is as hot as $T_{\rm SBN} \approx10^8$ K. 
Therefore, the sound speed of the free expanding wind at the shock is $c_s \approx 10^2 \rm \, T_6^{1/2} \, km \, s^{-1}$. 
As a consequence the Mach number of the plasma at the wind shock is of order $\sim 10$ making it the only plausible site for particle acceleration in the wind bubble system.

For the innermost regions of the system (the SBN and the cool wind) we adopt a smooth parametrization of the model of CC85 for the velocity profile (see bottom panel Figure~\ref{fig:Image}).
The model describes a wind where the velocity increases toward the edge of the SBN.  
At the SBN boundary the wind becomes supersonic and quickly reaches $V_{\infty}$, while beyond the termination shock, the gas gets heated and slowed down. 
At the termination shock we adopt the jump condition appropriate for a strong shock so that $u_1=V_{\infty}$ and $u_2=u_1/4$.
Moreover, for adiabatic expansion, the shocked wind moves with a velocity that drops with distance as $\sim r^{-2}$, namely $ur^2=$constant. The wind plasma is assumed to be fully ionized, while the density in the SBN is assumed to be dominated by dense molecular gas. Hence the particle density in the system (blue dot-dot-dashed curve in Figure~\ref{fig:Image}) can be approximated as:

\begin{equation}
    \rho(r) = \begin{cases}
        \rho_{\rm SBN} \hspace{1.5cm} r < R_{\rm SBN} \\
        \frac{\Dot{M}}{4 \pi r^2 v_w(r)} \hspace{0.8cm} R_{\rm SBN} < r < R_{\rm sh}  \\
        4 \times \frac{\Dot{M}}{4 \pi R_{\rm sh}^2 u_1} \hspace{0.65cm} R_{\rm sh} < r < R_{\rm FS}.
    \end{cases}
    \label{Eq: density}
\end{equation}
For the purpose of estimating the diffusion coefficient for high energy particles in the bubble, we assume that a fraction $\epsilon_{\rm B}$ (in MBPC21 we have used $\eta_B = \epsilon_B/2$) of the kinetic energy density of the free expanding wind is converted at any given radius into turbulent magnetic field energy density. We also assume that at the termination shock the perpendicular components of the magnetic field are compressed by a factor 4, which implies that the strength of the magnetic field downstream is enhanced by a factor $\sqrt{11}$ and remains spatially constant in the downstream region.

Overall, the strength of the magnetic field can be written as:
\begin{equation}
    B(r) = \begin{cases}
    \frac{\sqrt{\epsilon_{\rm B} \Dot{M} v_w(r) }}{r}  \qquad r < R_{\rm sh}  \\
    \sqrt{11} \times \frac{\sqrt{\epsilon_{\rm B} \Dot{M} u_1 }}{R_{\rm sh}}  \qquad r > R_{\rm sh}
    \end{cases}
     ,
    \label{Eq: magnetic-field}
\end{equation}
where, the radial dependence of the upstream wind profile $v_{w}(r)$, has a negligible impact on the magnetic field in the corresponding region.

Assuming that the turbulent field gets organized according to a power spectrum $P(k)\propto k^{-\delta}$, the corresponding diffusion coefficient due to resonant particle scattering can be estimated as:
\begin{equation}
    D(r,p) = \frac{1}{3} r_{\rm L}(r,p) \, v(p) \Big[\frac{L_c}{r_{\rm L}(r,p)} \Big]^{\delta-1} 
    \label{eq: Diff-Coeff}
\end{equation}
where $r_{\rm L}$ is the Larmor radius, $v$ the particle velocity and $\delta=5/3$ ($3/2$) for Kolmogorov (Kraichnan) turbulence. Bohm diffusion would correspond to $\delta=1$. 
The quantity $L_c$ denotes the energy containing scale of the turbulence. For momenta $p>p^*$, where $p^*$ is defined such that $r_L(p^*)=L_c$, the diffusion coefficient changes its energy dependence due to lesser power on larger scales and can be written as \cite[]{Subedi2017,Dundovic2020}:
\begin{equation}
    D(r,p) = \frac{1}{3} L_c \, v(p) \Big[ \frac{r_{\rm L}}{L_c} \Big]^2 \, \, \, \, p > p^* .
\end{equation}
In this work we adopt a Kraichnan spectrum of the turbulence, $\delta=3/2$ as the reference scenario and we assume $L_c$ to be comparable with the size of the SBN, namely $L_c \sim 10^2 \, \rm pc$.

\section{Model}
\label{Sec3: Model}

In this section we provide a detailed description of the theoretical model. 
In \S~\ref{SubS31: acc}, we present the solution of the CR transport equation of particles accelerated at the wind shock of the starburst--driven wind bubble. Together with the solution we additionally describe the flux of escaping particles.
In \S~\ref{Subs: Secondaries} we describe the calculation of gamma rays and neutrinos from pp and p$\gamma$ interactions.

\subsection{Particle acceleration at the termination shock}
\label{SubS31: acc}

Particle acceleration is assumed to take place at the termination shock. For the sake of simplicity we adopt a spherical symmetry neglecting the deformation induced by the surrounding medium.
The bubble is assumed to be already evolved through the deceleration phase, so that the shock location is given by Equation~\eqref{Eq: shock-radius}. 

The transport of non--thermal particles in the bubble is determined by diffusion, adiabatic energy losses and gains, advection with the wind and catastrophic energy losses, that are dominated by pp inelastic collisions in the SBN, for those particles that have high enough energy to diffuse against the wind and reach the central region. The transport equation that we solve can be written as follows:
\begin{eqnarray}
 \frac{\partial}{\partial r} \left[ r^2 D(r,p) \frac{\partial f}{\partial r} \right]
  - r^2 v_w(r) \frac{\partial f}{\partial r} 
  + \frac{d\left[ r^2 v_w(r) \right]}{d r} \frac{p}{3} \frac{\partial f}{\partial p} +  \nonumber \\
  + \, r^2 Q(r,p) - \, r^2 \Lambda(r,p) f = 0 .
  \label{Eq: transport}
\end{eqnarray}
where $f=f(r,p)$ is the particle distribution function, $D(r,p)$ is the diffusion coefficient (in general space dependent), $v_w(r)$ is the wind profile, $Q(r,p)$ is the injection term and $\Lambda(r,p)$ is the rate of energy losses.

Assuming that particle injection only takes place at the location of the termination shock and is limited to a single momentum $p_{\rm inj}$, we can write:
\begin{equation}
    \label{Eq: injection}
    Q(r,p) = \frac{Q_0(p)}{4\pi r^2} \, \delta[r-R_{\rm sh}] = \frac{\eta_{\rm inj} n_1 u_1}{4 \pi p^2} \, \delta[p-p_{\rm inj}] \, \delta[r-R_{\rm sh}],
\end{equation}
where $n_1$ and $u_1$ are the density and wind speed immediately upstream of the shock, and $\eta_{\rm inj}$ is the fraction of particles involved in the acceleration process. 
We take $\eta_{\rm inj}$ such that the pressure of accelerated particles is limited to a fraction, $\sim 10 \%$ of the wind ram pressure at the shock.
Notice that, as long as the shock compression factor is larger than $2.5$ (meaning that the spectrum is harder than $p^{-5}$), the value of $p_{\rm inj}$ does not play any relevant role in the normalization of Equation~\eqref{Eq: injection}.
The loss term takes into account energy losses for proton--proton collisions:
\begin{equation}
    \label{Eq: loss-term}
    \Lambda(r,p) = n(r) \sigma_{\rm pp}(p) v(p)  \, ,
\end{equation}
where $v$ is the particle speed, 
$n(r)=\rho(r)/m_p$ is the target density in the wind and $\sigma_{\rm pp}$ is the cross section \cite[]{Kelner_Aharonian_2006_proton-proton}. 
We neglect losses due to $p\gamma$ interactions since, as we show below, the maximum energy that particles reach is barely enough to exceed the kinematic threshold for this process, using optical (OPT) and ultraviolet (UV) photons as targets.

Equation \eqref{Eq: transport} is solved by following the technical procedure put forward in MBPC21 for the case of winds from star clusters. We refer the reader to that paper for details, while here we only summarize the main equations that allow us to obtain the solution of the problem by iterations. We also discuss the differences with respect to MBPC21, mainly due to the presence of energy losses.

The method starts from determining the solution of the transport equation upstream and downstream separately and then impose the continuity of the solution at the shock location.
The solution in the upstream region reads:
\begin{equation}
    \label{Eq: Upstream-solution}
    f_1(r,p) = f_{\rm sh}(p) \, e^{- \int_r^{R_{\rm sh}} d r' \, V_1(r',p)/D_1(r',p)},
\end{equation}
where $f_{\rm sh}$ is the particle distribution function at the shock and $V_1$ is an effective velocity \textit{felt} by particles upstream, due to the combination of spherical symmetry and energy losses:
\begin{equation}
    \label{Eq: V1_compount}
    V_1(r,p) = u_1(r) + \frac{G_1(r,p)+H_1(r,p)}{r^2 \, f_1(r,p)}.
\end{equation}
The functions $G_1$ and $H_1$  describe adiabatic energy losses--gains and catastrophic energy losses, respectively, and are reported in Appendix~\ref{Appendix: Calculations}

In the downstream region, the solution is made easier by the fact that the flow is divergence--free (namely $ur^2$ is constant) and energy losses due to pp scatterings are negligible.
This simplification allows us to write 
\begin{equation}
    \label{Eq: Downstream-solution}
    f_2(r,p) = f_{\rm sh}(p) \frac{1 - e^{\alpha(r,p)-\alpha(R_{\rm esc},p)}}{1-e^{-\alpha(R_{\rm esc},p)}}
\end{equation}
where
\begin{equation}
    \label{Eq: Alpha-down}
    \alpha(r,p) = \frac{R_{\rm sh} \, u_2}{D_2(p)} \left( 1 - \frac{R_{\rm sh}}{r} \right) 
\end{equation}
and $R_{\rm esc} \approx R_{\rm FS}$ is the location where particles escape from the system and assumed to be equal to the forward shock radius.
Integrating the transport equation in a narrow region around the termination shock we find an equation for $f_{\rm sh}(p)$, after using the solution upstream and downstream to evaluate the spatial derivatives on the two sides of the shock.
\begin{equation}
    \label{Eq: Shock-solution}
    f_{\rm sh}(p) = \frac{s \, n_1 \, \eta_{\rm CR} }{4 \, \pi \, p_{\rm inj}^3 } \, \left( \frac{p_{\rm inj}}{p} \right)^s \, e^{-\left[ \Gamma_1(p) + \Gamma_2(p) \right]}.
\end{equation}
Here $\Gamma_1$ and $\Gamma_2$ describe the departure from the standard solution $p^{-s}$ that would have been obtained at a plane infinite shock, due to a variety of factors:
\begin{eqnarray}
\label{Eq: E_max-GAMMA1}
    \Gamma_1(p) = s \int_{p_{\rm inj},}^{p} \frac{dp'}{p'} \frac{G_1(R_{\rm sh},p')+H_1(R_{\rm sh},p')}{u_1 R_s^2 f_{\rm sh}(p')} \,  \\ 
    \Gamma_2(p) = s \int_{p_{\rm inj},}^{p} \frac{dp'}{p'} \frac{u_2/u_1}{e^{\alpha(p',R_{\rm esc})} - 1} .
\label{Eq: E_max-GAMMA2}    
\end{eqnarray}
The function $\Gamma_1$ reflects the effects of spherical symmetry and losses upstream, and is appreciably different from unity at energies close to the maximum energy, namely at energies where $D_1/u_1$ becomes comparable to $R_{\rm sh}$ \citep[see also][]{Berezhko_1997}. 
For the particles that are energetic enough to reach the SBN, energy losses in the dense gas become important both for CR transport (if $\Dot{M}$ is large enough) and production of secondary radiation 
\citep[see][for related discussions]{Bustard_Zweibel2017,Merten_2018}.
However, in all cases that we have studied this phenomenon never leads to observable consequences.

Notice that Equation~\eqref{Eq: Shock-solution} expresses the solution in a recursive form, because both $G_1$ and $H_1$ are function of $f$. The actual solution is obtained using an iterative technique as described in MBPC21.

The spectral modification due to the transport in the downstream region is contained in the function $\Gamma_2$, which becomes important when the diffusion length of particles ($\lambda_D \sim D_2/u_2$) becomes comparable to the size of the shocked wind region ($R_{\rm esc}-R_{\rm sh}$). 

The escape flux at the bubble boundary, defined as $j_{\rm esc}= - D \, \partial_r f(R_{\rm esc})$, can be easily derived from Equation~\eqref{Eq: Downstream-solution}:
\begin{equation}
    j_{\rm esc}(p) = u_2 \, f_{\rm sh}(p) \frac{[R_{\rm sh}/R_{\rm esc}]^2}{1-e^{-\alpha(R_{\rm esc},p)}}.
\end{equation}
and the total flux of escaping particles is $J_{\rm esc} = 4 \pi R_{\rm esc}^2 \, j_{\rm esc}$.

The escape flux modifies the solution at the shock only very mildly and only for very high particle energies. On the other hand the spatial extent of the downstream region (shocked wind), which in turn depends on the age of the bubble, reflects rather strongly on the gamma--ray and neutrino signal from a SBG.

The assumption of stationarity adopted in the   equation requires that the acceleration process is much faster than the time for dynamical evolution of the system. This is typically the case, but as a consistency check, we always verify that the acceleration time defined as:
\begin{equation}
    t_{\rm acc}(p)= \frac{3}{u_1 - u_2} \int_0^p \frac{dp'}{p'} \left[ \frac{D_1(p')}{u_1} + \frac{D_2(p')}{u_2} \right]
\end{equation}
be shorter than the lifetime of the system \citep[see e.g.,][]{Blasi-review2013}.

\subsection{Production of secondaries}
\label{Subs: Secondaries}

As discussed in \S~\ref{SubS31: acc} (see also Appendix~\ref{Appendix: Bubble-lumin} for additional details), 
$pp$ and $p \gamma$ interactions in the downstream region take place with typical timescales larger than Gyr, so that their dynamical impact on the CR transport can be neglected.
However, the luminosity of the wind bubble can be a sizable fraction of the SBN's luminosity due to the large spatial extent of the system \citep[see also][for related discussions]{Romero-Muller,Muller-Romero}.

We thus compute the gamma--ray and neutrino emission resulting from the interaction of CRs with {\it i}) particles in the plasma through pp interactions and {\it ii}) thermal photons, as produced by stars and dust in the galaxy and illuminating the wind bubble itself ($p\gamma$ interactions).

The calculation of gamma--rays produced through pp interactions has been performed using the NAIMA package~\citep{kafexhiu2014} which implements the procedure described in \cite[]{Kelner_Aharonian_2006_proton-proton} while the gamma rays produced through $p \gamma$ interactions are computed following \cite[]{Kelner-photomeson}.
The gamma--ray absorption inside the SBN is taken into account as in~\cite[]{Peretti-1}, where the background photon field is assumed to be constant in the SBN volume.
On the other hand, the size of the system and the $r^{-2}$ dependence of the photon field imply negligible absorption effects for gamma rays produced in the wind bubble.
Finally, the gamma--ray absorption on the EBL on cosmological distances is computed adopting the EBL model of \citet{Franceschini-EBL}.

The single flavor neutrino flux is computed assuming equipartition among flavors, $(\nu_{e},\nu_{\mu},\nu_{\tau})=(1:1:1)$, due to flavor oscillations during propagation to the Earth. 
The production of neutrinos in pp interactions is estimated by rescaling the gamma--ray luminosity as: $L_{\nu}(E_{\nu}) \approx  L_{\gamma}(E_{\gamma})/2$, where $E_{\gamma} \approx 2 E_{\nu}$. The neutrinos produced in the $p \gamma$ interactions are computed following \citet{Kelner-photomeson}.

\section{Emission from individual starbursts}
\label{Sec4: Results - isolated galaxy}

In this section we discuss the results of the calculation of the spectra of accelerated particles and high energy gamma rays and neutrinos  (\S\ref{SubS: isolated source}) for an individual SBG and how the properties of the bubble and of the accelerated particles change when changing parameters (\S\ref{SubS: single-param-exp}).

\subsection{Particles and spectra}
\label{SubS: isolated source}

We discuss two stereotypical models of SBGs so as to illustrate how the results change by changing the properties of the SBN. The two benchmark cases are labelled as B0 and B1 and correspond to the parameters' values reported in Tab.~\ref{table:parameters-1}. 
The B0 prototype is reminiscent of local mild SBGs such as M82 and NGC253. We assume the photon field of NGC253~\citep{Galliano2008} as representative of the prototype B0. Observations and numerical simulations of M82 suggest a terminal (wind) velocity $ \sim 2000 \rm \, km \, s^{-1}$ \citep[][]{Strickland-Heckmann2009,Elizabete_2013}, with a mass loss rate up to $\lesssim 3 \, \rm M_{\odot} \, yr^{-1}$. Similar terminal wind speed but higher mass loss rate are inferred for NGC253 \citep[][]{Strickland_2002,Bolatto2013}.
The B1 configuration represents a somewhat more powerful wind that can be expected in objects for which the nuclear activity and temperature is higher (such as LIRGs) than what is inferred for M82 and NGC253~\citep[see e.g.,][]{Bustard_Zweibel2017}.
For the B1 prototype we assume that the photon background is somewhat larger than B0 and for reference we assume the SED of NGC1068~\citep{Galliano2008}.

In Table~\ref{table:parameters-1} we also show the maximum energy of accelerated particles, $E_{\max}$, and the single flavor neutrino energy flux produced in the wind bubble at 25~TeV, defined as ${\Tilde{F}_{\nu_{\mu}}}$, as observed from a distance of 3.9 Mpc. 
Both these quantities are outputs of our calculations. 
The positions of the termination shock, Equation~\eqref{Eq: shock-radius}, and of the edge of the bubble, Equation~\eqref{Eq: escape-radius}, for the two prototypes are calculated fixing the age of the system to $t^*_{\rm age} = 250 \, \rm Myr$ and by assuming a value of the pressure in the external medium, $P_{h}/k_B$.
\begin{table}
\caption{Parameters for the benchmark cases B0 and B1. The resulting maximum energy and monochromatic single flavor neutrino flux at 25 TeV are reported at the bottom. The latter is computed assuming a fiducial luminosity distance $d_L = 3.9 \, \rm Mpc$.}
\label{table:parameters-1}
\centering             
\linespread{1.15}\selectfont
\begin{tabular}{c|c|c}
\hline   
 & B0 & B1  \\
\hline  
$ \Dot{M} \rm [M_{\odot} \, yr^{-1}]$ & $5$ & $10$ \\
\hline 	
$V_{\infty}{\rm [10^3 \, km \, s^{-1}]}$ & $2$ & $3$   \\
\hline 
$P_{h}/k_B {\rm  [10^4 \, K \, cm^{-3}]}$ & $2.5$ & $8$  \\
\hline 
$t^*_{\rm age}{[\rm Myr]}$ & $250$ & $250$  \\
\hline 
$R_{\rm sh}{[\rm kpc]} $ & $12.36$ & $11.97$  \\
\hline 
$R_{\rm FS}/R_{\rm sh} $ & $4$ & $4.6$  \\
\hline \hline 
$E_{\rm max}{ \rm [PeV]}$ & $44$ & $131$  \\
\hline    
{$\Tilde{F}_{\nu_{\mu}}$}${\rm [10^{-11} \frac{GeV}{cm^2 \, s}]}$ & $1.27$ & $8.9$  \\
\hline                   
\end{tabular}
\linespread{1.0}\selectfont
\end{table}

\begin{figure}
	\includegraphics[width=\columnwidth]{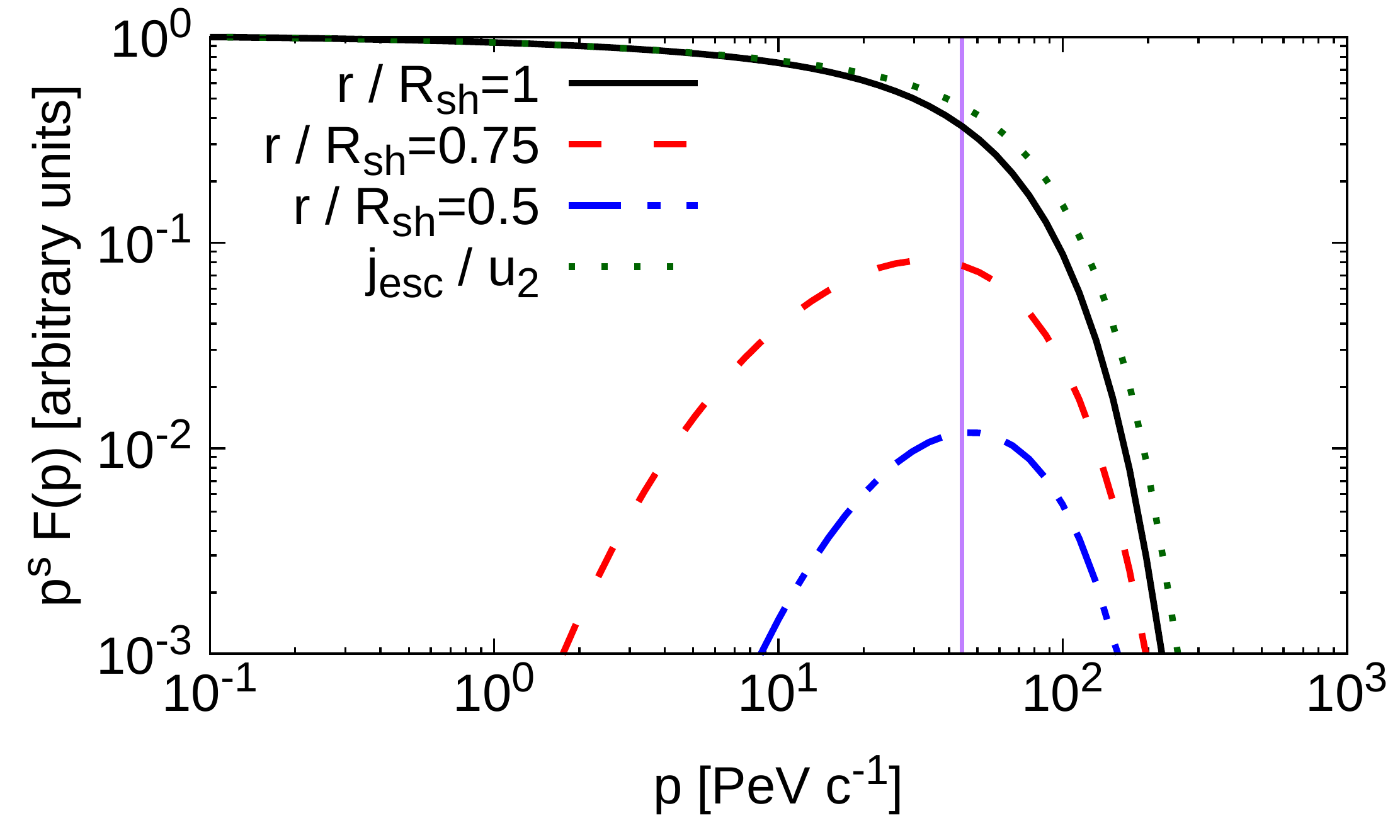} \quad \includegraphics[width=\columnwidth]{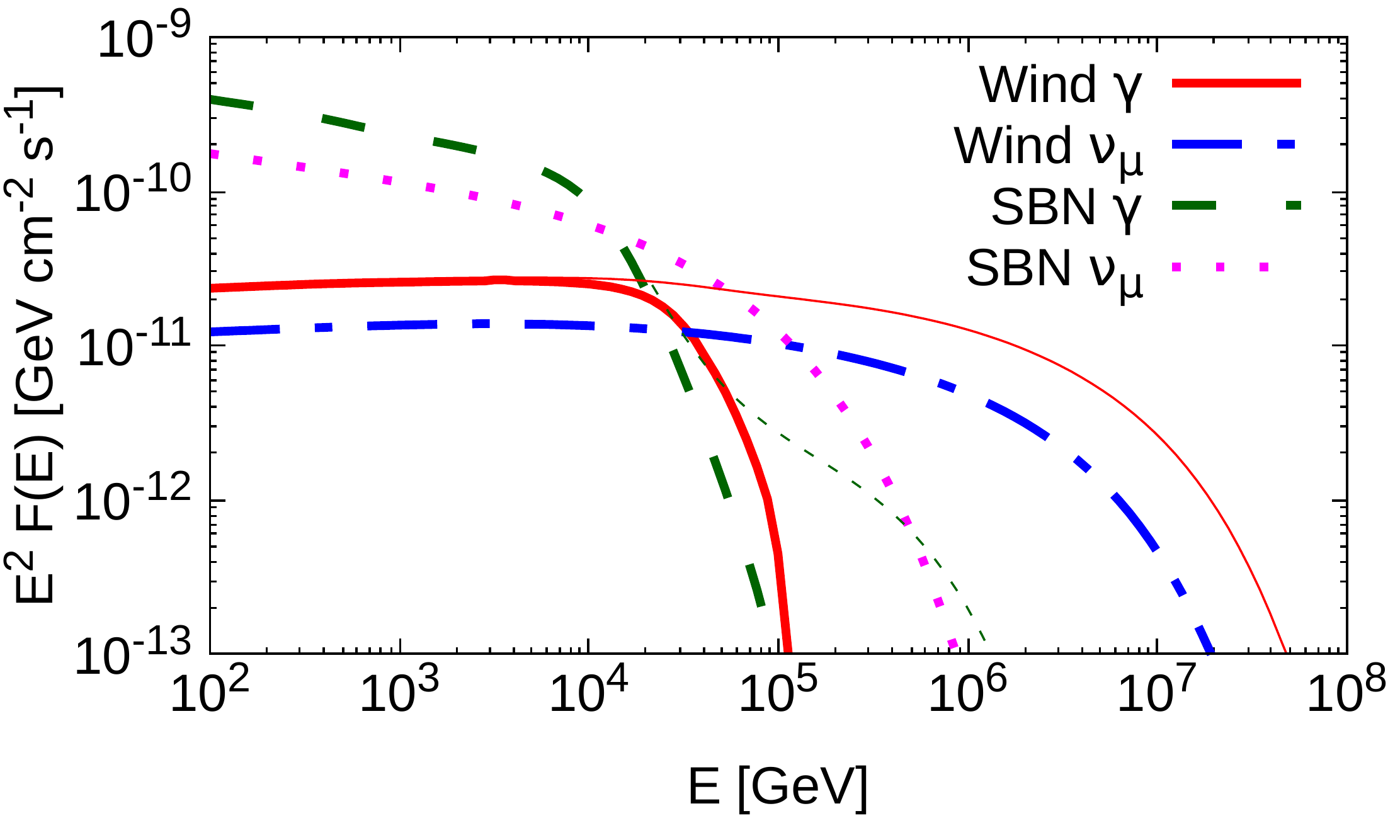}
    \caption{Particle spectrum and HE multimessenger spectra at Earth assuming $d_L= 3.9 \, \rm Mpc$ for the benchmark prototype B0. Top panel: proton spectrum at the shock (thick black line) compared to the solution at $0.75 \, R_{\rm sh}$ (red dashed line) and $0.5 \, R_{\rm sh}$ (blue dot--dashed line). The escape flux is also shown (green dotted line). Bottom panel: gamma--ray and neutrino flux from the wind (thick red and dot--dashed blue lines) compared to the emission from the SBN core (green dashed and pink dotted lines).
    The effect of EBL absorption is taken into account assuming a distance of $3.9$ Mpc. 
    For comparison, and for a qualitative view of the $\gamma \gamma$ absorption in the source, the gamma--ray components are shown when the EBL absorption is neglected (thin lines).}
    \label{fig:Image2}
\end{figure}

\begin{figure}
	\includegraphics[width=\columnwidth]{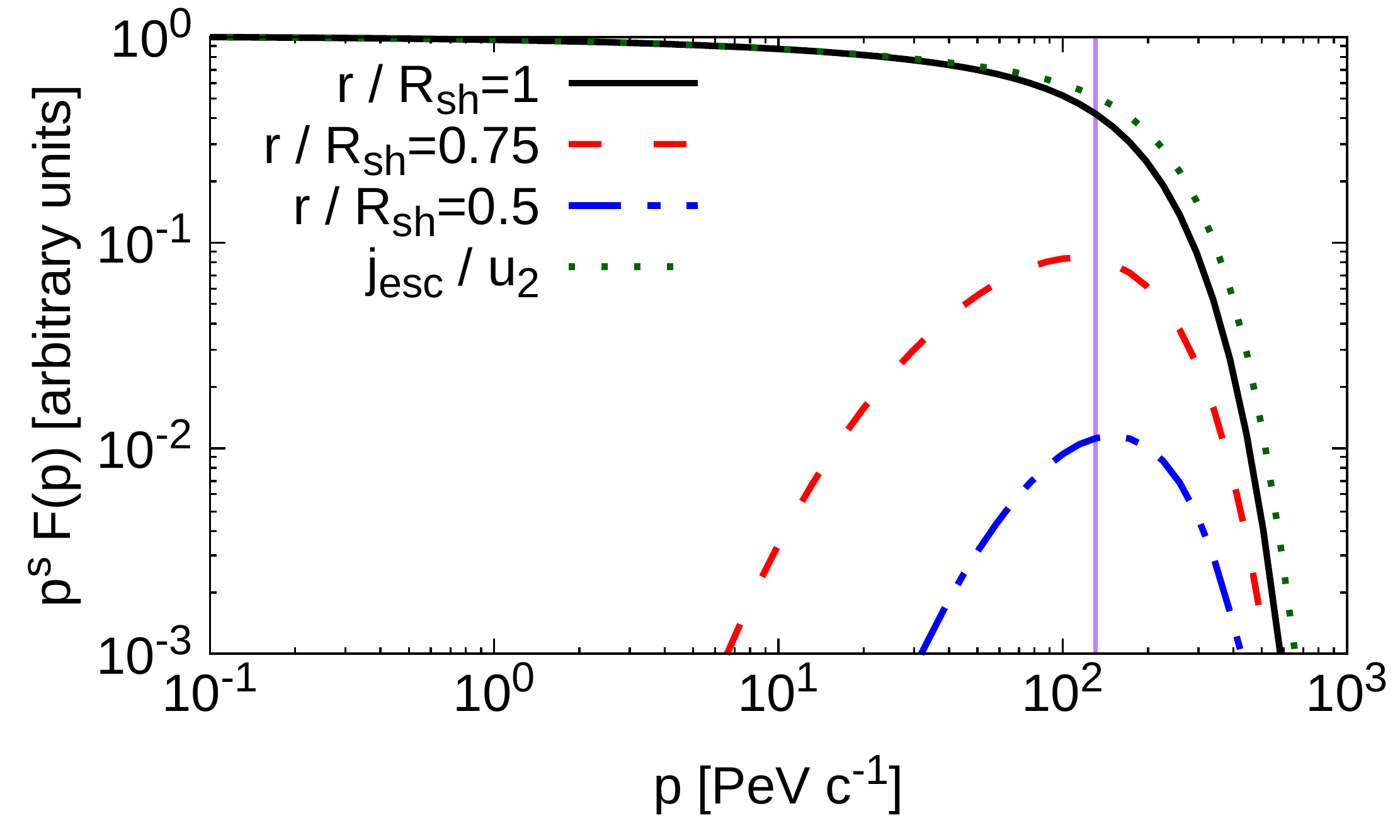} \quad \includegraphics[width=\columnwidth]{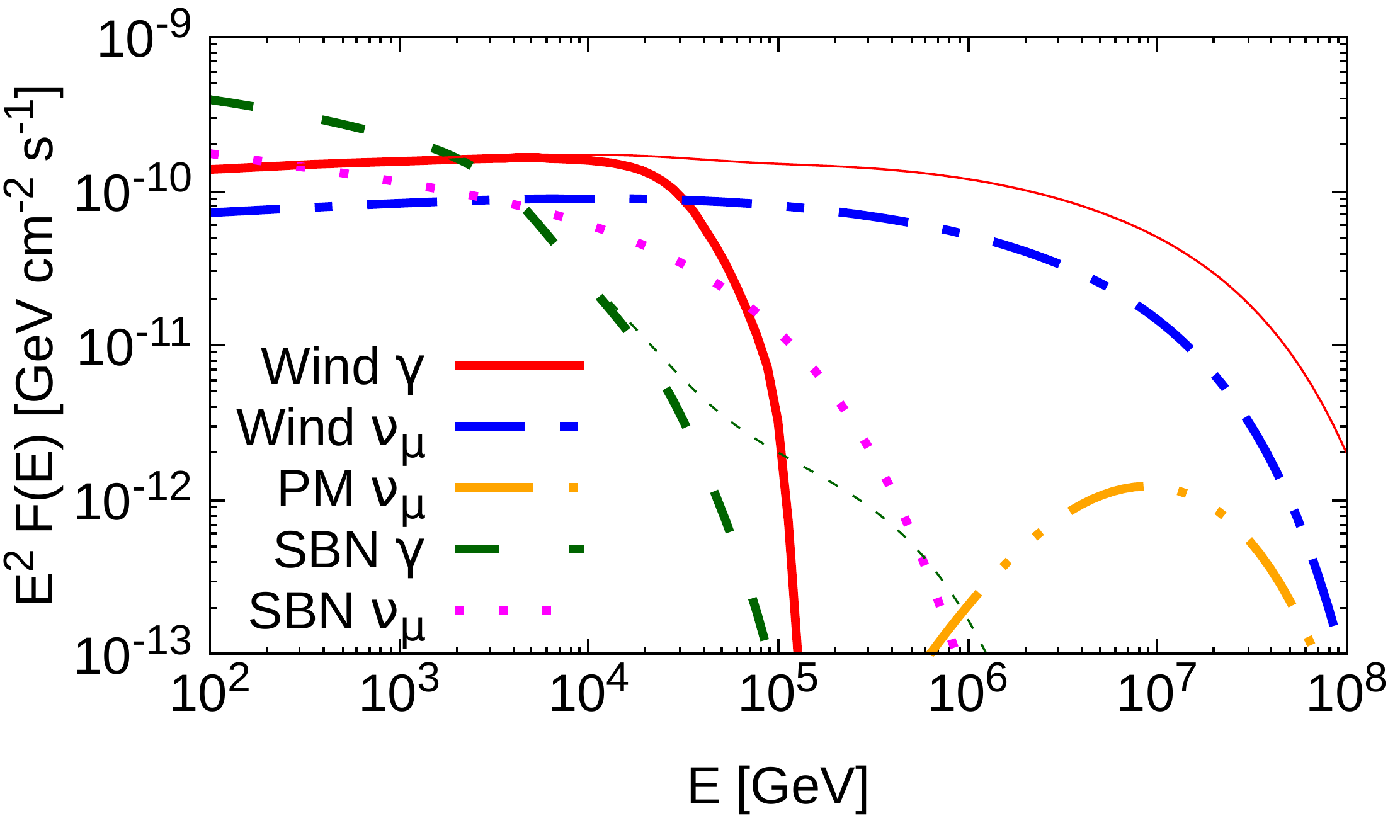}
    \caption{Particle and HE multimessenger spectra for the benchmark prototype B1. The line--style and colors are identical to Figure~\ref{fig:Image2}. The only difference is that the production of photomeson neutrinos from the wind bubble (orange dot--dot--dashed curve) becomes relevant at $\sim 10 \, \rm PeV$.}
    \label{fig:Image3}
\end{figure}

Results for the cases B0 and B1 are reported in Figures~\ref{fig:Image2} and \ref{fig:Image3}, respectively.
The top panels show the particle spectrum at the shock, the escaping flux and the particle spectrum in the cold wind region as computed at different radii ($0.75 R_{\rm sh}$ and $0.50  R_{\rm sh}$).
The vertical purple line identifies the position of the maximum momentum $p_{\rm max}$ of accelerated particles, defined as the value at which the spectrum $p^s f(p)$ is reduced by $e$.
The bottom panels of the same Figures show the corresponding spectra of gamma rays and neutrinos resulting from pp and p$\gamma$ interactions computed for the cases of a strong shock ($s=4$) and assuming that the source is located at a distance of $\sim 3.9 \, \rm Mpc$ (appropriate for M82). The red thick (thin) solid line shows the gamma emission from the wind region after (before) correcting for absorption on the EBL during transport from the source to Earth. 
Notice that the same plots report also the contribution of gamma rays and neutrinos produced by the interaction of CRs accelerated by SNRs inside the SBN and interacting inside the nucleus, assuming a source spectrum $\sim p^{-4.2}$, as inferred for M82 by \cite{Peretti-1}, with a maximum energy $\lesssim 1$ PeV. The thick (thin) line refers to the flux after (before) correction for absorption en route. 

The flux of muon neutrinos from the wind region is shown as a blue thick dash-dotted line. Such flux is dominated by the contribution of pion production in pp interactions downstream of the termination shock. In Figure~\ref{fig:Image3}, due to the larger luminosity of the SBG, the contribution to the neutrino flux due to photomeson production (dash dot-dotted orange line) becomes visible in the plot. Such flux is present only in the highest energy region because of the kinematic threshold of the process of photopion production.

A few comments on the spectrum of accelerated particles (top panels in Figure~\ref{fig:Image2} and \ref{fig:Image3}) are in order: as it would be the case for standard DSA, the spectrum of accelerated particles is a power law when the momentum is much smaller than the maximum one. On the other hand, as discussed in MBPC21, spherical symmetry induces a dependence of the spectrum on the diffusion coefficient that is most marked around $p_{\rm max}$. This is because particles can {\it feel} an effective plasma velocity which is smaller than $v_w$ when their diffusion length becomes comparable with $R_{\rm sh}$. Particles with high energies can travel farther away from the shock and feel its curvature in a more prominent way. The deviation from the standard power-law is more visible for weak energy dependence of the diffusion coefficient. In other words, the deviation from a power-law would start at lower energies for Kolmogorov diffusion, while it would occur closer to $p_{\rm max}$ for Bohm diffusion (see discussion in MBPC21). These subtle effects also reflect in the spectrum of secondary gamma rays and neutrinos. 
The maximum energy reached by accelerated particles varies between tens of PeV for the prototype B0 to $\gtrsim 100$ PeV for B1. 

As previously mentioned, here we assumed $s=4$, but in \S~\ref{Sec6: Discussion + Conclusions} we discuss the case of softer slopes as might arise due to the motion of scattering centers in the downstream plasma \citep[see][for details]{Caprioli2020}.

By looking at the particle spectrum in the inner region, one can conclude that only particles at the maximum energy can diffuse efficiently against the wind and populate the inner region of the system. 
Nevertheless, it appears clear that, unless an additional acceleration mechanism is present in the system, the number of particles that can successfully diffuse back to the SBN is strongly suppressed due to the geometry of the system. 
Indeed, in order to successfully diffuse upwind towards the SBN, particles need to have a momentum $p_{\rm b}$ such that the diffusion length becomes larger than the upstream region, namely $D(p_{\rm b})/u_1 \gtrsim R_{\rm sh}$. 
Finally, we notice that the spectrum of the escaping flux, as also discussed in MBPC21, does not differ strongly from the solution at the shock in terms of spectral slope and maximum energy.

The gamma--ray emission from the SBG is dominated by the emission of the SBN for $E\lesssim \rm TeV$. However, depending on the total power of the system and the conditions of the external medium where the bubble is expanding into, the emission from the wind region may become dominant at high enough energy and be identifiable as an extension of the spectrum up to the energy for which there is a substantial absorption on the EBL. 
In the scenario where accelerators in the SBN cannot exceed $\sim$PeV, all neutrinos with energy $\gtrsim 10^2 \, \rm TeV$ are produced in the wind and the luminosity increases with $\Dot{M}$ since this parameter directly affects the target density for pp interactions (see also Tab.~\ref{table:parameters-1}). 
The slope of the neutrino spectra below $\sim 10 \, \rm TeV$ is slightly harder than $E^{-2}$ due to the energy dependence of the cross section for pp inelastic collisions, $\sigma_{pp}$. Above $\sim 10 \, \rm TeV$ the spectral slope gets gradually softer due to the shape of the parent proton population.
The hadronic emission from the wind is dominated by the pp interaction taking place in the shocked wind region whereas the contribution from the free wind region might be relevant only for extreme values of $\Dot{M}$, or possibly during some early stages of the bubble evolution.
The photomeson contribution is found to be always subdominant compared to the pp and is irrelevant if $E_{\rm max} \ll 10^2 \, \rm PeV$, because of the kinematic threshold for this channel.
Finally, for some massive winds characterized by $\Dot{M} \gg 10 \rm \, M_{\odot} \, yr^{-1}$, the gamma--ray emission from the wind might become comparable with the SBN component even below the TeV range. 

\subsection{Exploring the parameter space}
\label{SubS: single-param-exp}

\begin{figure}
	\includegraphics[width=\columnwidth]{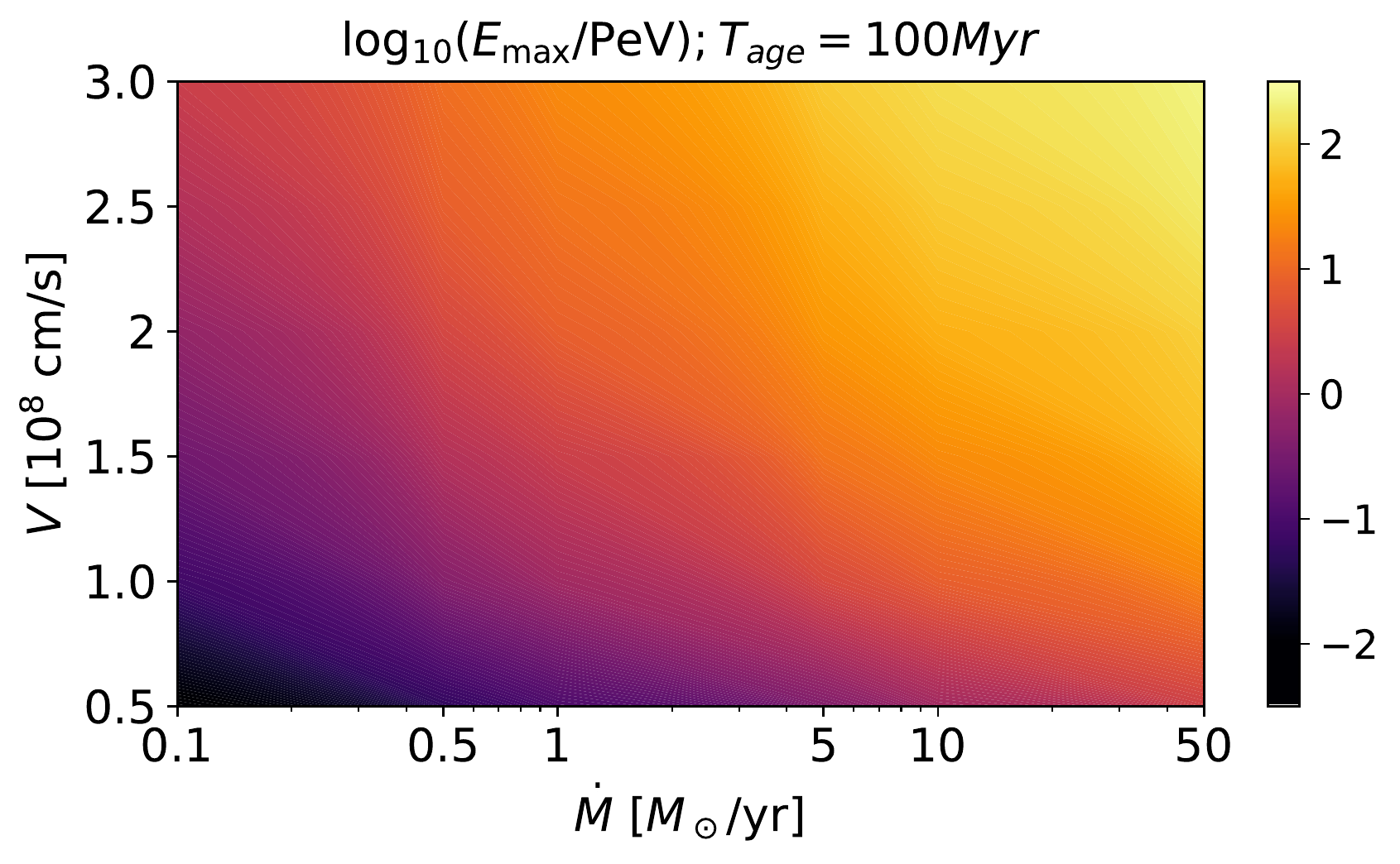} \quad \includegraphics[width=\columnwidth]{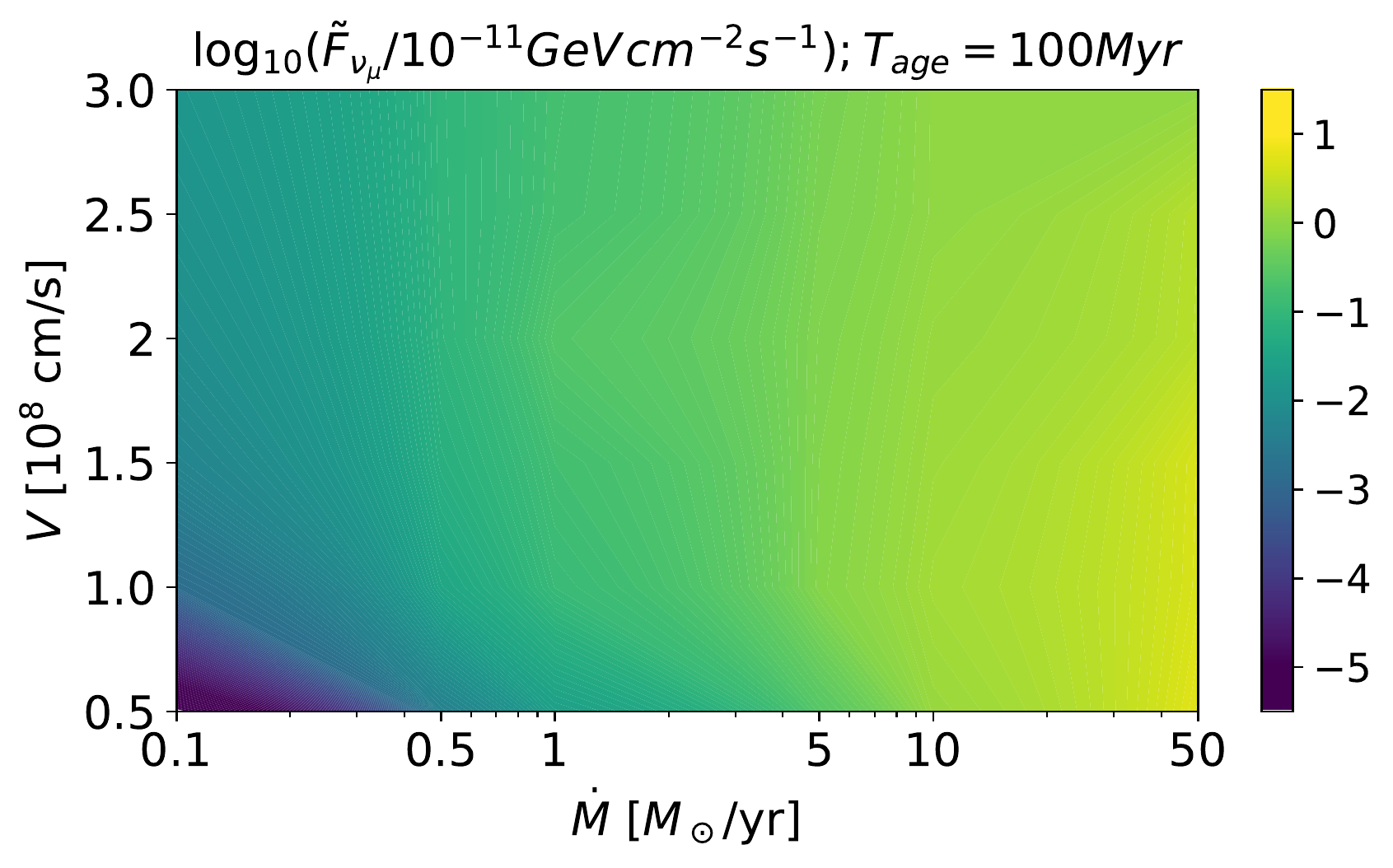}
    \caption{Contour plots illustrating the parameter space ($\Dot{M},V_{\infty}$) exploration based on different realizations of the system at the age $T_{\rm age,1}= 100$ Myr. The associated color code is shown at the right of each panel. Top panel: Maximum energy normalized to 1 PeV. Bottom panel: Single flavor neutrino flux at 25 TeV.}
    \label{fig:P100}
\end{figure}

\begin{figure}
	\includegraphics[width=\columnwidth]{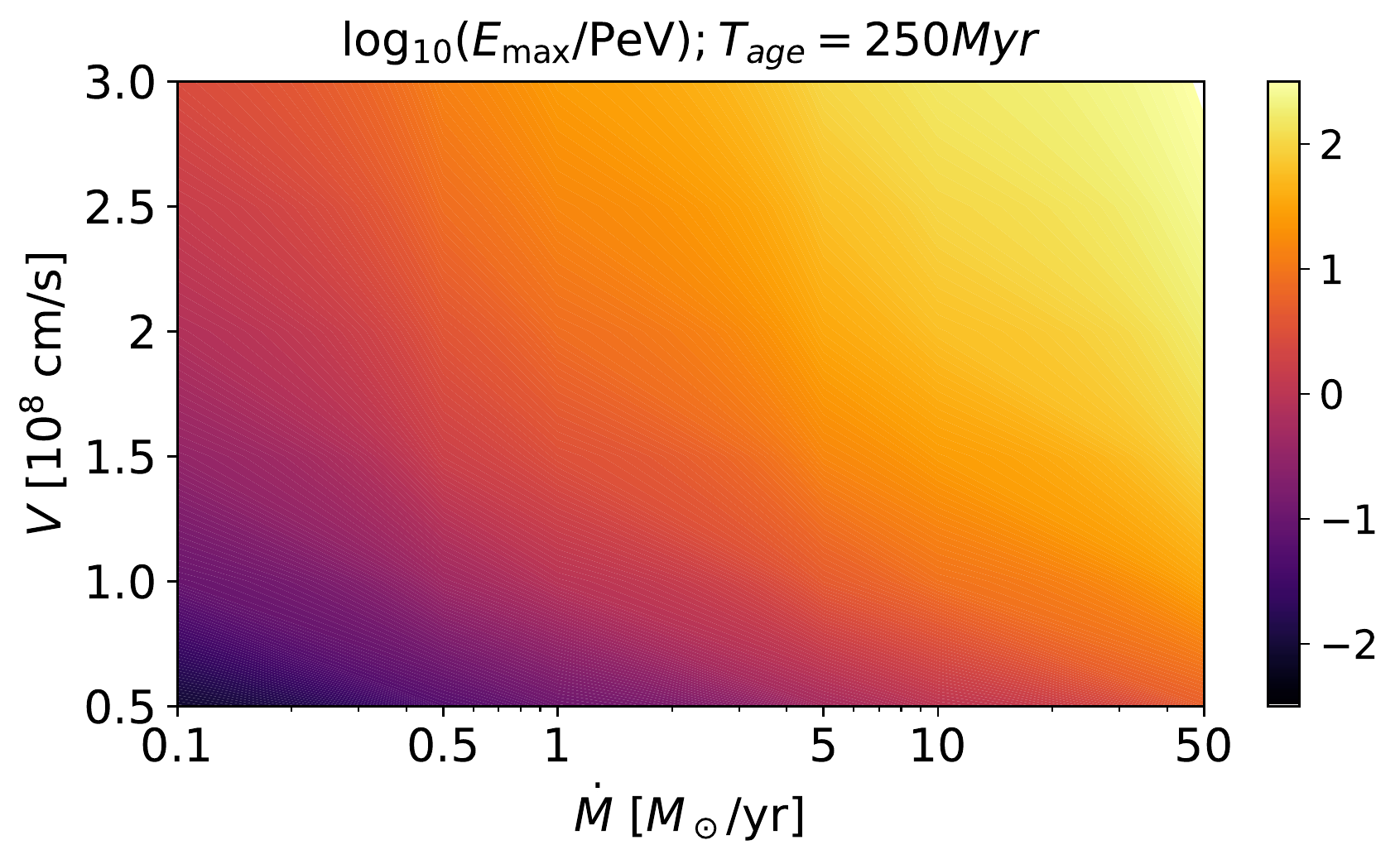} \quad \includegraphics[width=\columnwidth]{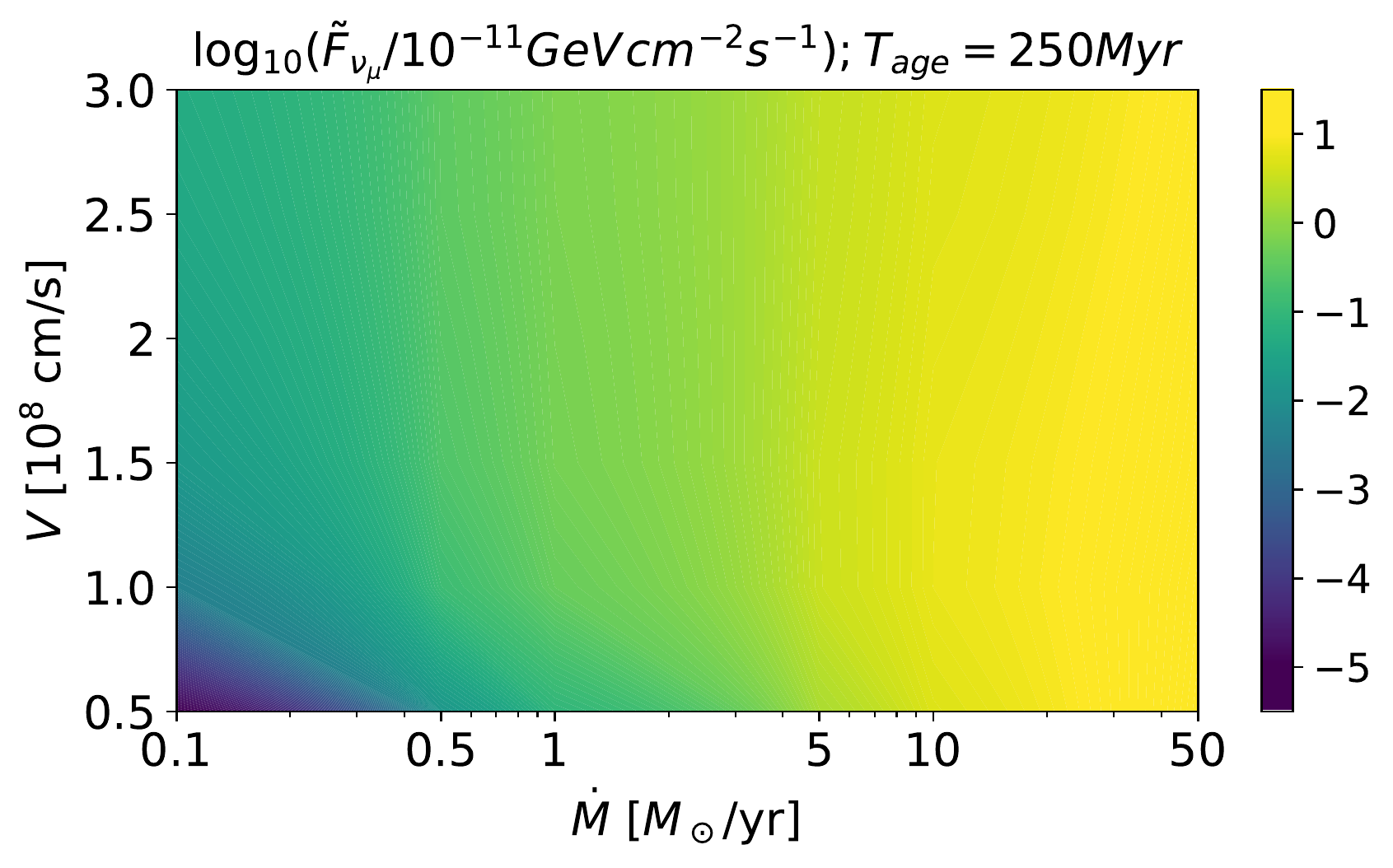}
    \caption{Contour plots illustrating the parameter space ($\Dot{M},V_{\infty}$) exploration based on different realizations of the system at the age $T_{\rm age,2}= 250$ Myr. Top and bottom panels are the same as Fig.~\ref{fig:P100}.}
    \label{fig:P250}
\end{figure}

In the discussion above we identified two main prototypical examples of SBGs, but clearly the zoo of these astrophysical objects cannot be reduced to just two cases. 
Here we provide a brief overview of what is expected to happen in different realizations of such systems. 
We do so by exploring a grid of different configurations of the main macroscopic wind properties, mass-loss rate ($\Dot{M}$) and terminal wind speed ($V_{\infty}$), and later by focusing on some specific parameter variations and the associated outcome. 
The corresponding relevant quantities are summarized in two pairs of plots (Figures~\ref{fig:P100} and \ref{fig:P250}) and in Table~\ref{table:parameters-2}. 
In what follows we focus on the effects of different conditions on: 1) maximum energy and 2) luminosity.

In our parameter investigation we define a range for the mass-loss rate, $0.1 <\Dot{M}/[M_{\odot} \, \rm yr^{-1}]< 50$ and for the terminal wind speed, $0.5<V_{\infty}/[10^3 \, \rm km \, s^{-1} ]<3$. 
In order to keep track of the temporal evolution, we additionally select two characteristic times at which we take a snapshot of the system: $T_{\rm age,1}= 100$ Myr and $T_{\rm age,2}= 250$ Myr.
In Figures~\ref{fig:P100} and \ref{fig:P250} we show the results obtained  at $T_{\rm age,1}$ and $T_{\rm age,2}$, respectively, under the assumption of $P_{h}/k_{B}= 5 \cdot 10^4 \, \rm K \, cm^{-3}$. 
The upper panels illustrate the changes in the maximum energy $E_{\rm max}$ and the lower panels show the single flavor neutrino flux at 25 TeV, $\Tilde{F}_{\nu_{\mu}}$.
%
%
In general, it can be observed that the higher the power of the system ($\Dot{M}V_{\infty}^2$), the higher the maximum energy. 
In particular, as discussed in \S\ref{Sec3: Model} \citep[see also][for additional details]{MBPC2021}, the most stringent condition on the maximum energy is typically set by the transport in the upstream region as $D(E_{\rm max}^{(1)})= R_{\rm sh} V_{\infty}$. 
Such a conditions can be re-expressed as
\begin{equation}
    E_{\rm max}^{(1)} = \Dot{M}^{1/[2(2-\delta)]} V_{\infty}^{3/[2(2-\delta)]} \frac{q_e \, \epsilon_B^{1/2} \, [3 \, l_c^{(1-\delta)} \, c^{-1} ]^{1/(2-\delta)} }{(4 \pi P_h)^{(\delta-1)/[2(2-\delta)]}},
    \label{Eq: E_max-proxi}
\end{equation}
which leads to $E_{\rm max}^{(1)} \propto \Dot{M} V_{\infty}^{3}$ for the assumed Kraichnan-like turbulence.
Notice that although the maximum energy in Eq.~\eqref{Eq: E_max-proxi} identifies an energy where the flux drops most prominently, as discussed above, the spherical symmetry of the problem leads to a gradual spectral steepening at energies below $E_{\rm max}$. This effect is embedded in the two functions $\Gamma_1$ and $\Gamma_2$ described in \S~\ref{Sec3: Model}.
At odds with the case of the maximum energy, the neutrino luminosity has a very mild dependence on the terminal wind speed provided that it is above the threshold to accelerate efficiently >PeV particles, whereas it has approximately a linear dependence on the mass-loss rate. The latter scaling is due to the direct connection between $\Dot{M}$ and the target density.

By comparing the results obtained at $T_{\rm age,1}$ with those obtained at $T_{\rm age,2}$ we observe that the age of the system does not have a dominant impact on the maximum energy as expected from Eq.~\eqref{Eq: E_max-proxi}: the acceleration time is much shorter than the dynamical time of these systems.
The slight difference can be understood by the interplay between the two functions, $\Gamma_1$ and $\Gamma_2$ regulating the HE cut-off. 
In fact, an older system is characterized by a less stringent constraint produced by $\Gamma_2$, while the one set by $\Gamma_1$ is practically unmodified.
On the other hand the luminosity is found to increase with time due to the increase of target material accumulated in the downstream region.

In order to evaluate the impact of changing other relevant parameters' values, we now focus on a set of limited cases listed in Table~\ref{table:parameters-2} and discuss quantitatively their numerical outcomes.
We first change the total luminosity of the system: L1 corresponds to a strong wind as the one that can be found in LIRGs; L2 corresponds to a mild star forming source. 
In line with what we discussed above, these two situations illustrate that, maintaining the same halo conditions, the maximum energy increases with the power of the wind. 
The location of the wind termination shock as well as of the forward shock is moved farther away from the center when the power is larger. %
Consequently, the most powerful sources naturally lead to a larger volume of the bubble and higher gamma--ray and neutrino luminosity.

In scenarios P1, P2 and P3, the total power is as in B0, but the surrounding pressure in the halo varies by 3 orders of magnitude. 
This again impacts the location of the termination shock $R_{\rm sh}$ which in turn affects the maximum energy even though this latter quantity varies only by a factor of $\sim 2$. 
In particular, in agreement with Eq.~\eqref{Eq: E_max-proxi}, the smaller the halo pressure, the higher the maximum energy . 
Although Eq.~\eqref{Eq: E_max-proxi} is informative on the dependence of the maximum energy on the CGM pressure, the actual scaling of $E_{\rm max}$ on $P_h$ is not straightforward due to the role played by the functions $\Gamma_1$ and $\Gamma_2$ in shaping the spectrum close to $E_{\rm max}$, and due to the transition of the diffusion coefficient to the $\sim E^2$ regime, when $r_{\rm L}\approx L_c$. The last effect is in fact occurring at energies close to the actual $E_{\rm max}$.
Scenario P3 corresponds to a relatively extreme situation since the wind evolves in a very low pressure environment compared to what might be expected in a starburst halo. Under these conditions, the system would need $\sim 5 \, \rm Gyr$ to reach the pressure--confined state. Consequently, both the forward and the wind shocks are still in their expansion phase after 250 Myr. 
In this case, the wind shock radius cannot be computed under the pressure confined assumption, so that we adopt equation~(4.2) of \citet[][]{Koo-McKee}.
In this scenario the maximum energy is somewhat close to the scenario labelled as P2, but the luminosity is smaller due to the lower ram pressure at the shock which, in turn, results from the larger shock distance from the center. 
Even if the impact on the maximum energy is marginal, the value of the circumgalactic pressure strongly impacts on the gamma-ray and neutrino luminosity. This is a direct consequence of the assumed proportionality between the CR energy density and the free wind ram pressure which is, in turn, roughly equal to the external pressure. Comparing cases P1 with P3, where $P_h$ is $10^3$ smaller, the neutrino luminosity decreases by $\sim 140$ times. The proportionality is not exactly linear because the spatial distribution of both CRs and gas in the two cases is different.

By comparing scenario L2 with P1 and L1 with P2, one can notice that sources with similar age and size can strongly differ both in maximum energy and luminosity. 
The former result can be easily understood given the dependence of the maximum energy on the $V_{\infty}$ (see Equations~\eqref{Eq: E_max-GAMMA1}, \eqref{Eq: E_max-GAMMA2} and Eq.~\eqref{Eq: E_max-proxi}). 
The luminosity, on the other hand, is set by the combination of the pressure at the shock, which determines the total number of accelerated particles, and the target density (which in turn depends on $\Dot{M}$). 

\begin{table}
\caption{Different scenarios considered. The physical units are the same as in Tab.~\ref{table:parameters-1}. For the case P3  the wind bubble is not pressure confinement and different equations are used to estimate the bubble properties.}
\label{table:parameters-2}
\centering             
\linespread{1.15}\selectfont
\begin{tabular}{c|c|c|c|c|c|c|c}
\hline   
 & L1 & L2 & P1 & P2 & P3 & T1 & T2 \\
\hline  
$\Dot{M}$ & $20$ & $2$ & $5$ & $5$ & $5$ & $5$ & $5$  \\	
\hline 	
$V_{\infty}$ & $3$ & $1$ & $2$ & $2$ & $2$ & $2$ & $2$  \\
\hline 
$P_{h}/k $ & $2.5$ & $2.5$ & $10$ & $0.5$ & $0.01$ & $2.5$ & $2.5$  \\
\hline 
$t^*_{\rm age}$ & $250$ & $250$ &  $250$ & $250$ & $250$ & $100$ & $300$  \\
\hline 
$R_{\rm sh} $ & $30$ & $5.5$ & $6$ & $28$ & $40$ & $12$ & $12$  \\
\hline 
$R_{\rm FS}/R_{\rm sh} $ & $2.5$ & $5.5$ & $6.7$ & $2.5$ & $2$ & $2.2$ & $4.3$   \\
\hline \hline 
$E_{\rm max}$ & $225$ & $2$ & $26$ & $59$ & $58$ & $34$ & $44$  \\
\hline
{$\Tilde{F}_{\nu_{\mu}}$} & $2.6$ & $0.45$ & $8.7$ & $0.15$ & $0.06$ & $0.24$ & $1.5$  \\

\hline               
\end{tabular}
\linespread{1.0}\selectfont
\end{table}
Finally, T1 and T2 correspond to B0 at different times $t_{\rm age}$, 100 Myr and  300 Myr, illustrating that the slow evolution in time of the system does not have a strong impact on the maximum energy.
However, sources become more luminous while ageing, due to the target material that accumulates and larger volume of the shocked wind region where pp interactions are taking place.
This supports numerically what has been discussed above {based upon} the contour plots parameter investigation.

\section{Diffuse fluxes of cosmic rays, gamma rays and neutrinos}
\label{Sec5: Results - diffuse}

In this section we illustrate our calculations of the diffuse flux of gamma rays, neutrinos and cosmic rays due to the collective emission of starburst galactic winds distributed in redshift. 
In \S~\ref{SubS: Gamma+NU--diFF} we evaluate the starburst contribution to the diffuse fluxes of gamma rays and neutrinos and compare them to those observed by Fermi--LAT \citep[][]{Fermi-LAT<820} and IceCube \citep[][]{IceCube2020_LAST} respectively. 
In \S~\ref{SubS: CR--diFF} we explore the associated flux of CR protons accelerated at the termination shock and eventually escaping the bubble.

\subsection{Gamma rays and neutrinos}
\label{SubS: Gamma+NU--diFF}

We work under the assumption that starburst winds are ubiquitous in SBGs and we count sources following the star formation rate function (SFRF) approach previously adopted by \citet{Peretti-2} and defined for redshift up to $z=4.2$ \citep[see also][for additional details]{Gruppioni_2015}. 
Differently from the case of SBNi, where, as discussed by \citet[][]{Peretti-2}, the luminosity scales with the SFR, the dependence of the wind properties on the SFR is highly non trivial and difficult to constrain. Therefore, in the following we rely on the assumption that on average all winds above a given SFR value, $\psi_{\rm min}$, can be described in terms of a single prototype. The diffuse flux can be computed as
\begin{align}
    \Phi_{j}(E)= &  \frac{1}{4 \pi} \int d \Omega \int_0^{4.2} dz \; \frac{dV_{\rm C}{ (z) }}{dz \, d \Omega} \,  e^{-\tau_{j}(E,z)} \,\times \nonumber \\   
    & \int_{\psi_{\rm min}}^{\infty} d \log \psi \; \Phi_{\rm SFR}(\psi,{ z}) \; {[1+z]^2} f_{j}(E{ [1+z]},\psi),
    \label{Eq: diffuse-flux}
\end{align}
where $f_{j}(E,\psi)$ is the flux density of the particle specie $j=\{\gamma,\nu\}$, $\Phi_{\rm SFRF}$ is the SFRF, $dV_{\rm C}= c D_{\rm C}^2(z)/[E(z) \, H_0] \, dz \, d \Omega $ is the comoving volume element per redshift interval $dz$ and solid angle $d \Omega$. 
In a spatially flat space--time $D_{\rm C}(z)= D_{\rm L}(z)/(1+z)$ and $E(z)= \sqrt{\Omega_{\rm M} (1+z)^3 + \Omega_{\Lambda}}$.  
The quantity $\tau_{j}$ is assumed to vanish in the neutrino case while, in the case of gamma rays, it represents the opacity due to the presence of the EBL and cosmic microwave background (CMB) \cite[]{Franceschini-EBL}.
The contribution of the electromagnetic cascade is computed as in~\citet[][]{Peretti-2} \citep[see also][for additional details]{Berezinsky_em_cascade:2016}.
Finally, $\psi_{\rm min}$ represents the minimum star formation rate that we adopt as a free parameter considering the value of $\psi^* \sim 1 \, \rm M_{\odot} \, yr^{-1}$ \cite[]{Peretti-2} as a firm lower limit. 
The assumption of $\psi_{\rm min}$ as free parameter is dictated by the poorly constrained ratio between the mass--loss rate of the wind and the star formation rate in the SBN, $\mathcal{R}=\Dot{M}/SFR$ \citep[see e.g.,][for a detailed discussion]{Veilleux}. 
In general $\mathcal{R}\lesssim 2$ \citep[see also][for detailed discussions]{Bustard-2016,Zhang2018}, hence we fix $\mathcal{R} = 2$. 
Therefore, $\psi_{\rm min}$ increases with the wind mass--loss rate.

\begin{figure}
	\includegraphics[width=\columnwidth]{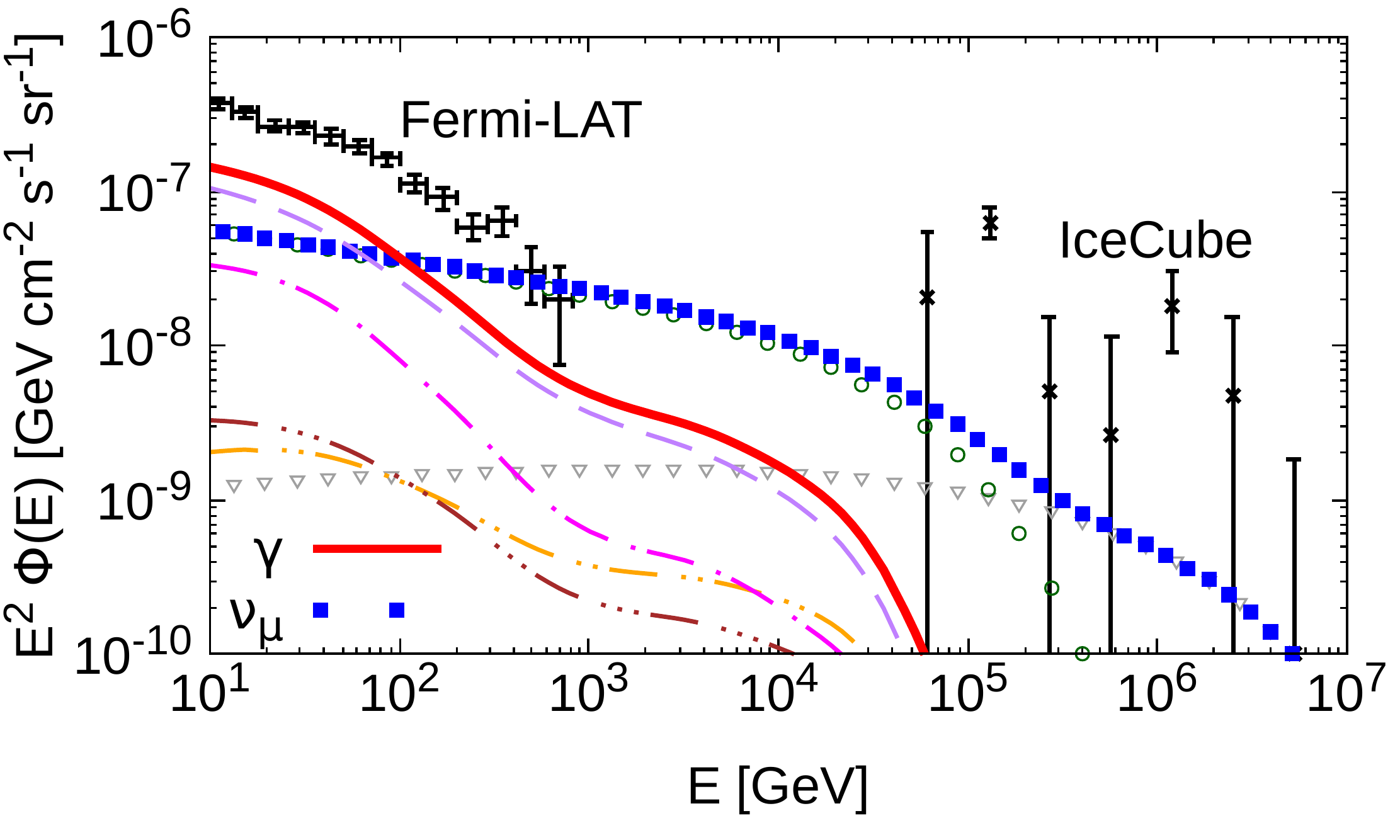} \quad \includegraphics[width=\columnwidth]{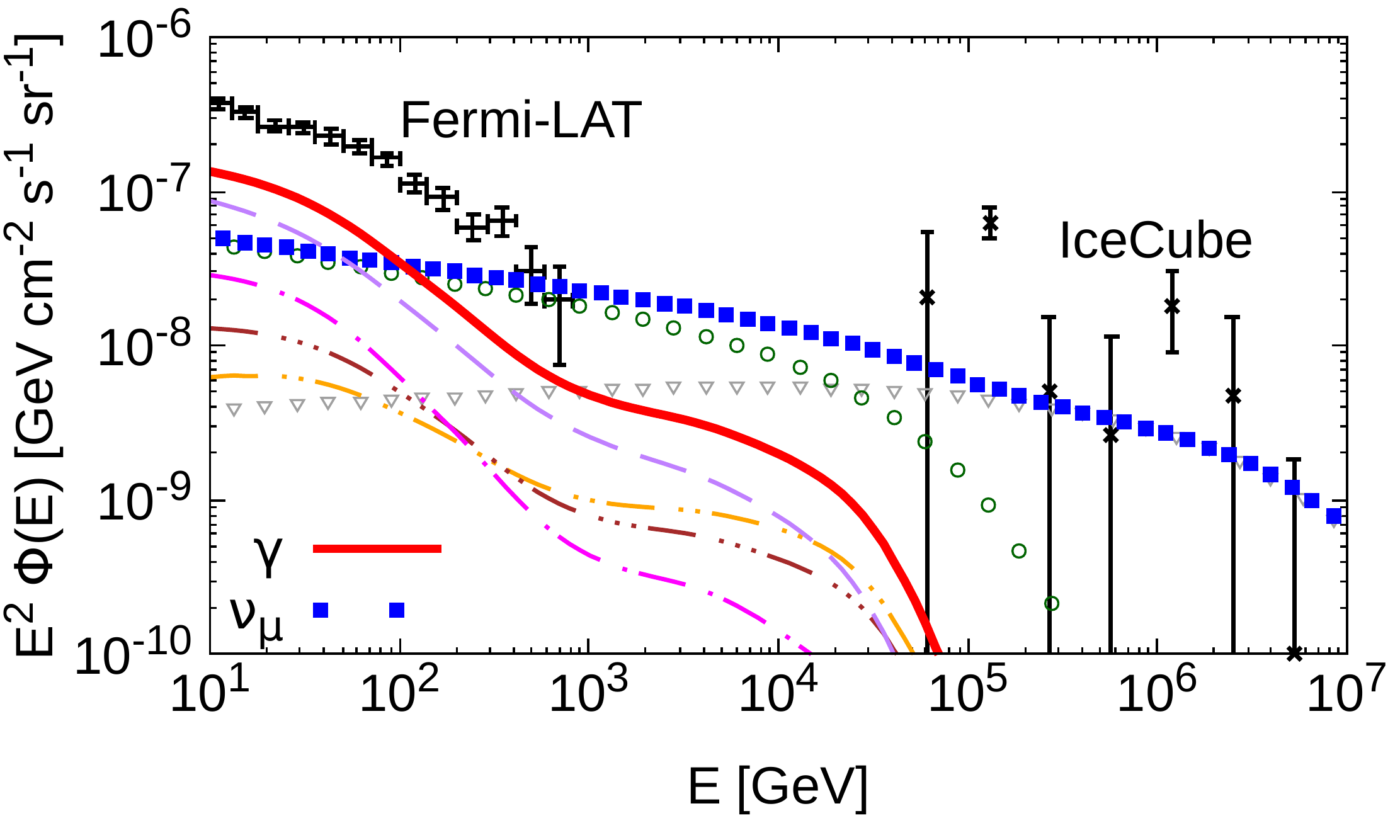}  
    \caption{Multimessenger emission for scenarios I and II (top and bottom panels, respectively) compared with Fermi--LAT \citep[][]{Fermi-LAT<820} and IceCube \citep[][]{IceCube2020_LAST} data. The color code is the same for all panels: total gamma rays and single flavor neutrinos are shown as thick red lines and blue filled squares respectively. Direct gamma--ray component from the SBN and wind (dashed violet and two--dot--dashed orange respectively) are shown separately with their associated cascade spectra (dot--dashed magenta and three--dot--dashed respectively). The neutrino components from SBNi (green empty circles) and from the winds (grey empty triangles) are shown separately.}
    \label{fig:Image4}
\end{figure}

We calculate the diffuse emission from the SBN and from its wind in two scenarios, that we refer to here as I and II, in which B0 and B1 are respectively used as a prototype (see Table~\ref{table:parameters-1}).
Following our criterion on $\mathcal{R}$, we adopt a $\psi_{\rm min}$ of 2.5 and 5 $\rm M_{\odot} \, yr^{-1}$ for cases I and II, respectively. 

In Figure~\ref{fig:Image4} we show the spectra of diffuse $\gamma$-rays and $\nu_{\mu}$ for the two scenarios I and II (top and bottom panels, respectively). 
In both cases the central SBN provides the main contribution to the total gamma-ray diffuse flux (dashed violet line and thick red line respectively). 
The latter lies below the diffuse flux measured by Fermi--LAT and never exceeds the upper limits imposed by the superposition of point--like sources \citep[e.g.][]{Lisanti_2016}. 
As described above, the wind region also contributes to the gamma--ray emission, and the corresponding diffuse flux is shown as an orange two--dot--dashed line in Figure~\ref{fig:Image4}. The cascade components (three--dot--dashed brown and dot--dashed magenta for SBNi and winds respectively) are always subdominant and change their relative contribution depending on the scenario.

The neutrino flux from SBNi (empty green circles) drops considerably above $\sim 50$ TeV, as a result of the proton maximum energy at sources in the SBN being $\sim 1$\,PeV. 
The flux of neutrinos produced in the wind (empty gray triangles) through pp collisions extends to $\gtrsim 300$ TeV and dominates the diffuse emission at such energies, at least at the level of $\sim 10^{-9} \, \rm GeV \, cm^{-2} \, s^{-1} \, sr^{-1}$ in the most pessimistic scenario (case I) \citep[this lower limit corresponds to $\sim 10 \%$ of the IceCube flux of through--going muons reported by ][]{IceCube_muons}.  
We finally notice that, if star forming galaxies were dominating the diffuse gamma-ray flux as suggested by \citet{Linden_2017,Roth-nature}, the associated neutrino flux would correspondingly increase.

\subsection{Cosmic rays}
\label{SubS: CR--diFF}

\begin{figure}
	\includegraphics[width=\columnwidth]{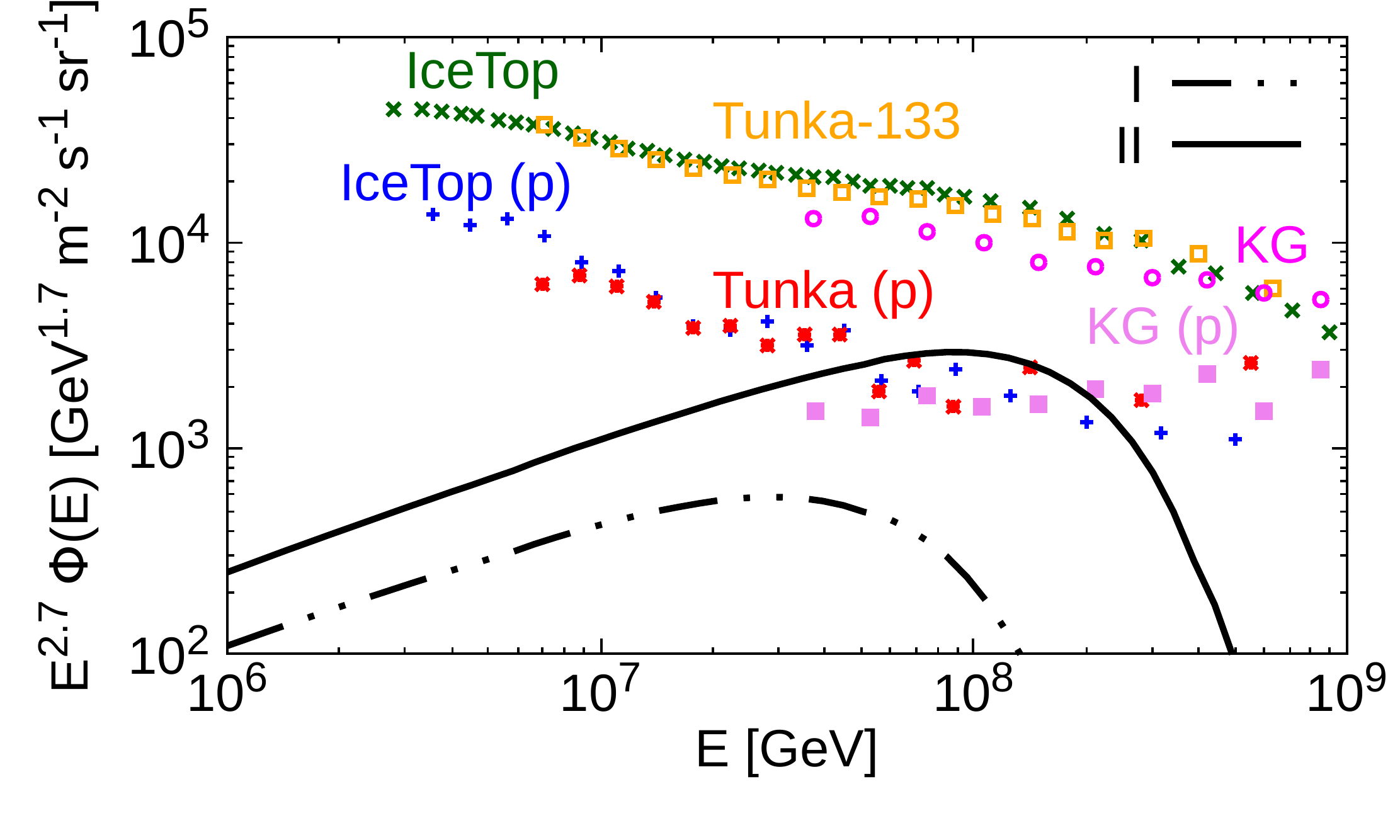}
    \caption{Diffuse proton flux escaped from the starburst wind and propagated to the Earth for the prototype cases B0 (dot-dot-dashed) and B1 (thick).  
    The predicted proton flux is compared with data from Tunka \citep[][]{Tunka2013,Tunka2016}, Kascade-GRANDE \citep[][]{Kascade-GRANDE2017} and IceTop \citep[][]{IceTop2019}.}
    \label{fig:Image5}
\end{figure}

CR protons accelerated at the termination shock of the SBG wind eventually escape the system from the outer edge of the bubble. Since energy losses do not affect the spectrum of these particles in a significant way, the escape spectrum is similar to the spectrum of particles accelerated at the termination shock. The diffuse flux of protons contributed by SBG winds, calculated using Equation~\eqref{Eq: diffuse-flux} which neglects any propagation effects due to the intergalactic magnetic fields, is shown in Figure~\ref{fig:Image5} for the scenarios I and II introduced earlier. Notice that since the maximum energy of accelerated particles is $\lesssim$ few hundred PeV, below the threshold for Bethe-Heitler pair production, the transport of these CRs on cosmological scales is dominated by adiabatic losses alone as due to the expansion of the Universe. In Figure~\ref{fig:Image5}, the predicted proton fluxes are compared with data of the all-particle spectrum as well as on the light component alone, as collected by IceTop \citep[][]{IceTop2019}, Tunka \citep[][]{Tunka2013,Tunka2016} and Kascade--Grande \citep[][]{Kascade-GRANDE2017}.
This shows that if indeed particle acceleration at winds termination shock does take place, so as to contribute to the high energy neutrino flux, a sizeable contribution to the protons CR flux measured at the Earth should be expected. 
Notice that here we only estimated the flux of protons from SBGs, but it is reasonable to expect that heavier nuclei are also accelerated, if present in the wind. Such nuclei would contribute to the total CR flux at higher energies.
We also observe that our results on the starburst contribution to the CR spectrum are qualitatively supported by \citet{Zhang+Murase2020} where, however, different assumptions were adopted for both the acceleration and transport of high-energy particles in galactic winds.

A comment on the spectral shape of CRs from SBGs is in order: one can see in Figure~\ref{fig:Image5} that the spectrum expected at the Earth is similar to that originated at individual wind bubbles, as a consequence of the fact that adiabatic losses do not change the spectral shape. On the other hand, such a straightforward connection can be made here only because of the assumption that all SBGs can be considered as similar to one of the two prototypical sources adopted here. In general this is not the case, and one should expect that the higher the wind luminosity the higher the maximum energy of the accelerated particles, but the lower the number of such objects in the Universe. 
As a result, qualitatively, one might expect that the diffuse flux of CR protons (as well as neutrinos) might become steeper at energies higher than the maximum energy associated to the least luminous of the winds, as discussed in a generic case by \cite{Kach}. 
We finally observe that, based on our calculations, it is difficult to accelerate protons above $\sim 10^{18}$ eV in the wind of normal SBGs.

\section{Discussion and conclusion}
\label{Sec6: Discussion + Conclusions}

The theory of particle acceleration at the termination shock of winds originating in star clusters, as developed by \citet{MBPC2021}, has been adapted here to the description of particle acceleration at the termination shock of starburst winds. %
At such shock the wind from the SBN is slowed down and heated up, so as to reach approximate pressure balance with the 
galactic halo in which the wind was originally expanding.
In fact a weak forward shock moves slowly through the 
halo medium, but its Mach number is too low to be of relevance for particle acceleration. 

We have assumed a stationary spherical geometry for the wind blown bubble. 
Even though numerical simulations might show a variety of possible deviations from such an assumption, particle acceleration and transport are not particularly affected by such details.
The theoretical approach is used to calculate the spectrum of accelerated particles and their spatial distribution inside the wind bubble, as well as their escape flux from the edge of the bubble. We discussed two prototypical SBG models, assumed to represent respectively a galaxy like M82 or NGC253 and a LIRG, and for each the flux of gamma rays and neutrinos produced due to CR interactions in the SBN and in the wind bubble has been calculated. 
The absorption of the gamma rays both inside the nucleus and {\it en route} to the Earth has been taken into account. 

The maximum energy of accelerated particles at the termination shock varies between a few tens PeV and 200 PeV for the two prototypes of SBGs considered here and in a range from a few PeV up to a few hundred PeV exploring a wider range of parameters. 
This implies that the corresponding neutrino flux extends up to 1-10 PeV, while the neutrino flux from the SBN is expected to extend up to a few tens of TeV, if CRs are accelerated by SNRs as in the Milky Way. Given the fact that the termination shock is strong, for the parameters adopted here, the spectrum of accelerated particles at $E\ll E_{\rm max}$ is close to $p^{-4}$. Some theoretical arguments can be put forward, for instance based on a finite velocity of scattering centers in the downstream region, to argue that slightly steeper spectra are possible ~\citep[see e.g.][]{Caprioli2020}.

The diffuse gamma--ray flux due to the superposition of SBGs is dominated by the contribution of the central SBNi for energies $\lesssim 1$ TeV. This flux does not exceed the upper limits imposed by Fermi-LAT based on the contribution of point--like sources \citep[e.g.][]{Lisanti_2016}. 
The wind region also contributes to the gamma--ray emission and such contribution can become comparable with that of the SBNi for $E\gtrsim 1$ TeV, if the more luminous prototype is adopted in the calculation of the diffuse flux.  
The neutrino flux from SBNi drops considerably above $\sim 50$ TeV, as a result of the proton maximum energy at sources in the SBN. On the contrary, the flux of neutrinos produced in the wind through pp collisions extends to $\gtrsim 300$ TeV and dominates the diffuse emission at such energies. The diffuse flux in this energy region is compatible with the IceCube data. 

The observational confirmation that particle acceleration at the termination shock and production of gamma rays and neutrinos in the wind bubble do take place can be achieved to some extent with upcoming observational facilities, as we discuss below. 
The study of starburst--driven galactic winds is generally performed via atomic and molecular line shifts and measurements of the X--ray luminosity \citep[][]{Veilleux,Strickland-Heckmann2009}, but so far, detection in the gamma--ray domain are rather limited, and unable to resolve the SBN emission from a possible contribution from the wind bubble.
A gamma--ray survey would be ideal to probe the model discussed in this work and would provide key information on its acceleration properties and luminosity. 
However, the most useful information would come from direct detection of the gamma--ray emission from the wind region. In the VHE range, the nearest starbursts, M82 and NGC253, could already be resolved by current instruments. In fact, a bubble of size $\sim 50 \rm \, kpc$, at a distance of $\sim 3-4 \, \rm Mpc$, corresponds to an angular size $\theta \sim 1^{\circ}$, typically resolved by  imaging air Cherenkov telescopes (IACTs) \citep[][]{Veritas-sensitivity,MAGIC-sensitivity,HESS}. However, given the total volume integrated luminosity of the order of $L_{\rm \gamma}(10 \, {\rm TeV}) \sim 10^{41} \, \rm GeV \, s^{-1}$ 
, expected for these sources, this task remains challenging.  

Next generation IACTs, such as ASTRI and CTA, with improved angular resolution and sensitivity, will open promising perspectives for a morphological study of these sources~\citep{Astri,CTA}.
A second method for probing the gamma--ray emission of the starburst wind consists in a spectral detection of the source at energy $E \gtrsim 10 \, \rm TeV$. The main reason for that is because gamma--gamma absorption on the IR is expected to be important above a few TeV in the SBNi. 
Differently, the emission from the wind comes basically unabsorbed. 
The observation of non--thermal radio/X--ray emission at large distances from the galactic disk can also be adopted to trace both the acceleration of primary electrons and the presence of secondaries produced via pp and p$\gamma$ interactions. A multiwavelength modeling focused on the leptonic emission is left for future investigation.

The detectability of SBGs as isolated neutrino sources is disfavored for the standard parameters adopted in this work. 
However, very young systems or scenarios involving high mass loss rates, $\Dot{M} \gtrsim 10 \, \rm M_{\odot} \, yr^{-1}$, can possibly produce fluxes close to the sensitivity level of km--squared detectors \citep[][]{KM3NeT,IceCube-Gen2}.
Differently from a single isolated source, the combined contribution of SBGs might provide interesting indications with higher statistical significance. 

Finally, we checked that the flux of CR protons accelerated at the termination shock and eventually propagating to the Earth is not in conflict with present day observation of the protons spectrum. 
In fact the diffuse flux of CRs from starburst wind bubbles is tantalizingly close to the observed flux, and limited to energies $\lesssim$ a few hundreds PeV, although it cannot be excluded that ultra--luminous SBGs or SBG with AGN activity may lead to the production of CRs with somewhat larger energies. 

However, the role of SBGs in contributing to the observed CR flux at $\sim 10^{17} \, \rm eV$ needs some additional support to be more robust than an order of magnitude estimation. 
Finally, in the context of the model developed in this work, we do not expect regular starbursts to be able to produce protons at energies larger than a few hundred PeV. We cannot exclude that higher energies may be reached in galaxies with SB activity hosting AGN jets, where particle acceleration would be regulated by different physical processes.

\section*{Acknowledgements}

This project has received funding from the European Union’s Horizon 2020 research and innovation program under the Marie Sklodowska-Curie grant agreement No. 847523 ‘INTERACTIONS’. The research activity of EP was partially supported by Villum Fonden (project n. 18994). The research activity of PB and GM was partially funded through support Grant ASI/INAF n. 2017-14-H.O; GM also was funded through Grants SKA-CTA-INAF 2016 and INAF-Mainstream 2018.
EP is grateful to M. Ahlers, A. Lamastra and E.M. de Gouveia Dal Pino for useful comments and to O. Pezzi, A. Dundovic, D. Boncioli, S. Petrera and A. Condorelli for helpful discussions at the initial stage of the work.

\section*{Data Availability}

No data has been analyzed or produced in this work. 
The phenomenological predictions performed in this work are compared with the data produced and analyzed in available publications. 
In particular, Fermi--LAT gamma--ray data can be found in \citet[][]{Fermi-LAT<820} and IceCube neutrino data are published in \citet[][]{IceCube2020_LAST}.
Cosmic--ray data measured by IceTop, Kascade--GRANDE and Tunka can be found respectively in \citet[][]{IceTop2019}, \citet[][]{Kascade-GRANDE2017} and \citet[][]{Tunka2013,Tunka2016}.



\bibliographystyle{mnras}
\bibliography{example} 

\begin{thebibliography}{}
\makeatletter
\relax
\def\mn@urlcharsother{\let\do\@makeother \do\$\do\&\do\#\do\^\do\_\do\%\do\~}
\def\mn@doi{\begingroup\mn@urlcharsother \@ifnextchar [ {\mn@doi@}
  {\mn@doi@[]}}
\def\mn@doi@[#1]#2{\def\@tempa{#1}\ifx\@tempa\@empty \href
  {http://dx.doi.org/#2} {doi:#2}\else \href {http://dx.doi.org/#2} {#1}\fi
  \endgroup}
\def\mn@eprint#1#2{\mn@eprint@#1:#2::\@nil}
\def\mn@eprint@arXiv#1{\href {http://arxiv.org/abs/#1} {{\tt arXiv:#1}}}
\def\mn@eprint@dblp#1{\href {http://dblp.uni-trier.de/rec/bibtex/#1.xml}
  {dblp:#1}}
\def\mn@eprint@#1:#2:#3:#4\@nil{\def\@tempa {#1}\def\@tempb {#2}\def\@tempc
  {#3}\ifx \@tempc \@empty \let \@tempc \@tempb \let \@tempb \@tempa \fi \ifx
  \@tempb \@empty \def\@tempb {arXiv}\fi \@ifundefined
  {mn@eprint@\@tempb}{\@tempb:\@tempc}{\expandafter \expandafter \csname
  mn@eprint@\@tempb\endcsname \expandafter{\@tempc}}}

\bibitem[\protect\citeauthoryear{{Aab} et~al.}{{Aab} et~al.}{2018}]{Auger-anis}
{Aab} A.,  et~al., 2018, \mn@doi [Astrophysical Journal Letters]
  {10.3847/2041-8213/aaa66d}, \href
  {https://ui.adsabs.harvard.edu/abs/2018ApJ...853L..29A} {853, L29}

\bibitem[\protect\citeauthoryear{{Aartsen} et~al.}{{Aartsen}
  et~al.}{2019}]{IceTop2019}
{Aartsen} M.~G.,  et~al., 2019, \mn@doi [\prd] {10.1103/PhysRevD.100.082002},
  \href {https://ui.adsabs.harvard.edu/abs/2019PhRvD.100h2002A} {100, 082002}

\bibitem[\protect\citeauthoryear{{Aartsen} et~al.}{{Aartsen}
  et~al.}{2021}]{IceCube-Gen2}
{Aartsen} M.~G.,  et~al., 2021, \mn@doi [Journal of Physics G Nuclear Physics]
  {10.1088/1361-6471/abbd48}, \href
  {https://ui.adsabs.harvard.edu/abs/2021JPhG...48f0501A} {48, 060501}

\bibitem[\protect\citeauthoryear{{Abbasi} et~al.}{{Abbasi}
  et~al.}{2020}]{IceCube2020_LAST}
{Abbasi} R.,  et~al., 2020, arXiv e-prints, \href
  {https://ui.adsabs.harvard.edu/abs/2020arXiv201103545A} {p. arXiv:2011.03545}

\bibitem[\protect\citeauthoryear{Abdalla et~al.}{Abdalla
  et~al.}{2018}]{Abdalla:2018nlz}
Abdalla H.,  et~al., 2018, Submitted to: Astron. Astrophys.

\bibitem[\protect\citeauthoryear{{Acharya} et~al.}{{Acharya}
  et~al.}{2019}]{CTA}
{Acharya} B.~S.,  et~al., 2019, {Science with the Cherenkov Telescope Array}.
World Scientific Publishing Co. Pte. Ltd., \mn@doi{10.1142/10986}

\bibitem[\protect\citeauthoryear{{Ackermann} et~al.}{{Ackermann}
  et~al.}{2012}]{Ackermann_Fermi_2012}
{Ackermann} M.,  et~al., 2012, \mn@doi [\apj] {10.1088/0004-637X/755/2/164},
  \href {http://adsabs.harvard.edu/abs/2012ApJ...755..164A} {755, 164}

\bibitem[\protect\citeauthoryear{Ackermann et~al.}{Ackermann
  et~al.}{2015}]{Fermi-LAT<820}
Ackermann M.,  et~al., 2015, \mn@doi [Astrophys. J.]
  {10.1088/0004-637X/799/1/86}, 799, 86

\bibitem[\protect\citeauthoryear{{Aiello}, others  \& {KM3NeT
  Collaboration}}{{Aiello} et~al.}{2019}]{KM3NeT}
{Aiello} S.,  others  {KM3NeT Collaboration} 2019, \mn@doi [Astroparticle
  Physics] {10.1016/j.astropartphys.2019.04.002}, \href
  {https://ui.adsabs.harvard.edu/abs/2019APh...111..100A} {111, 100}

\bibitem[\protect\citeauthoryear{{Ajello}, {Di Mauro}, {Paliya}  \&
  {Garrappa}}{{Ajello} et~al.}{2020}]{Ajello2020}
{Ajello} M.,  {Di Mauro} M.,  {Paliya} V.~S.,   {Garrappa} S.,  2020, \mn@doi
  [\apj] {10.3847/1538-4357/ab86a6}, \href
  {https://ui.adsabs.harvard.edu/abs/2020ApJ...894...88A} {894, 88}

\bibitem[\protect\citeauthoryear{{Aleksi{\'c}} et~al.}{{Aleksi{\'c}}
  et~al.}{2016}]{MAGIC-sensitivity}
{Aleksi{\'c}} J.,  et~al., 2016, \mn@doi [Astroparticle Physics]
  {10.1016/j.astropartphys.2015.02.005}, \href
  {https://ui.adsabs.harvard.edu/abs/2016APh....72...76A} {72, 76}

\bibitem[\protect\citeauthoryear{{Ambrosone}, {Chianese}, {Fiorillo},
  {Marinelli}, {Miele}  \& {Pisanti}}{{Ambrosone} et~al.}{2020}]{Ambrosone2020}
{Ambrosone} A.,  {Chianese} M.,  {Fiorillo} D. F.~G.,  {Marinelli} A.,  {Miele}
  G.,   {Pisanti} O.,  2020, arXiv e-prints, \href
  {https://ui.adsabs.harvard.edu/abs/2020arXiv201102483A} {p. arXiv:2011.02483}

\bibitem[\protect\citeauthoryear{{Ambrosone}, {Chianese}, {Fiorillo},
  {Marinelli}  \& {Miele}}{{Ambrosone} et~al.}{2021}]{Ambrosone2}
{Ambrosone} A.,  {Chianese} M.,  {Fiorillo} D. F.~G.,  {Marinelli} A.,
  {Miele} G.,  2021, \mn@doi [\apjl] {10.3847/2041-8213/ac25ff}, \href
  {https://ui.adsabs.harvard.edu/abs/2021ApJ...919L..32A} {919, L32}

\bibitem[\protect\citeauthoryear{{Anchordoqui}}{{Anchordoqui}}{2018}]{Anchordoqui2018}
{Anchordoqui} L.~A.,  2018, \mn@doi [\prd] {10.1103/PhysRevD.97.063010}, \href
  {https://ui.adsabs.harvard.edu/abs/2018PhRvD..97f3010A} {97, 063010}

\bibitem[\protect\citeauthoryear{Anchordoqui, Romero  \& Combi}{Anchordoqui
  et~al.}{1999}]{Anchordoqui-Romero1}
Anchordoqui L.~A.,  Romero G.~E.,   Combi J.~A.,  1999, \mn@doi [Phys. Rev. D]
  {10.1103/PhysRevD.60.103001}, 60, 103001

\bibitem[\protect\citeauthoryear{Anderson, Churazov  \& Bregman}{Anderson
  et~al.}{2015}]{Anderson2015-halos}
Anderson M.~E.,  Churazov E.,   Bregman J.~N.,  2015, \mn@doi [Monthly Notices
  of the Royal Astronomical Society] {10.1093/mnras/stv2314}, 455, 227

\bibitem[\protect\citeauthoryear{Arteaga-Vel\'azquez
  et~al.}{Arteaga-Vel\'azquez et~al.}{2018}]{Kascade-GRANDE2017}
Arteaga-Vel\'azquez C.~J.,  et~al., 2018, \mn@doi [PoS] {10.22323/1.301.0316},
  ICRC2017, 316

\bibitem[\protect\citeauthoryear{Bechtol, Ahlers, Di~Mauro, Ajello  \&
  Vandenbroucke}{Bechtol et~al.}{2017}]{Bechtol-Ahlers:2015}
Bechtol K.,  Ahlers M.,  Di~Mauro M.,  Ajello M.,   Vandenbroucke J.,  2017,
  \mn@doi [Astrophys. J.] {10.3847/1538-4357/836/1/47}, 836, 47

\bibitem[\protect\citeauthoryear{{Berezhko} \& {V{\"o}lk}}{{Berezhko} \&
  {V{\"o}lk}}{1997}]{Berezhko_1997}
{Berezhko} E.~G.,  {V{\"o}lk} H.~J.,  1997, \mn@doi [Astroparticle Physics]
  {10.1016/S0927-6505(97)00016-9}, 7, 183

\bibitem[\protect\citeauthoryear{Berezinsky \& Kalashev}{Berezinsky \&
  Kalashev}{2016}]{Berezinsky_em_cascade:2016}
Berezinsky V.,  Kalashev O.,  2016, \mn@doi [Phys. Rev.]
  {10.1103/PhysRevD.94.023007}, D94, 023007

\bibitem[\protect\citeauthoryear{{Blasi}}{{Blasi}}{2013}]{Blasi-review2013}
{Blasi} P.,  2013, \mn@doi [\aapr] {10.1007/s00159-013-0070-7}, \href
  {https://ui.adsabs.harvard.edu/abs/2013A&ARv..21...70B} {21, 70}

\bibitem[\protect\citeauthoryear{{Blasi} \& {Amato}}{{Blasi} \&
  {Amato}}{2019}]{Blasi-Amato2019}
{Blasi} P.,  {Amato} E.,  2019, \mn@doi [\prl]
  {10.1103/PhysRevLett.122.051101}, \href
  {https://ui.adsabs.harvard.edu/abs/2019PhRvL.122e1101B} {122, 051101}

\bibitem[\protect\citeauthoryear{{Bolatto} et~al.,}{{Bolatto}
  et~al.}{2013}]{Bolatto2013}
{Bolatto} A.~D.,  et~al., 2013, \mn@doi [\nat] {10.1038/nature12351}, \href
  {https://ui.adsabs.harvard.edu/abs/2013Natur.499..450B} {499, 450}

\bibitem[\protect\citeauthoryear{{Breitschwerdt}, {McKenzie}  \&
  {Voelk}}{{Breitschwerdt} et~al.}{1991}]{Breitshwerdt91}
{Breitschwerdt} D.,  {McKenzie} J.~F.,   {Voelk} H.~J.,  1991, \aap, \href
  {https://ui.adsabs.harvard.edu/abs/1991A&A...245...79B} {245, 79}

\bibitem[\protect\citeauthoryear{{Buckman}, {Linden}  \& {Thompson}}{{Buckman}
  et~al.}{2020}]{Buckman2020}
{Buckman} B.~J.,  {Linden} T.,   {Thompson} T.~A.,  2020, \mn@doi [\mnras]
  {10.1093/mnras/staa875}, \href
  {https://ui.adsabs.harvard.edu/abs/2020MNRAS.494.2679B} {494, 2679}

\bibitem[\protect\citeauthoryear{{Bustard}, {Zweibel}  \& {D'Onghia}}{{Bustard}
  et~al.}{2016}]{Bustard-2016}
{Bustard} C.,  {Zweibel} E.~G.,   {D'Onghia} E.,  2016, \mn@doi [\apj]
  {10.3847/0004-637X/819/1/29}, \href
  {https://ui.adsabs.harvard.edu/abs/2016ApJ...819...29B} {819, 29}

\bibitem[\protect\citeauthoryear{{Bustard}, {Zweibel}  \& {Cotter}}{{Bustard}
  et~al.}{2017}]{Bustard_Zweibel2017}
{Bustard} C.,  {Zweibel} E.~G.,   {Cotter} C.,  2017, \mn@doi [\apj]
  {10.3847/1538-4357/835/1/72}, \href
  {https://ui.adsabs.harvard.edu/abs/2017ApJ...835...72B} {835, 72}

\bibitem[\protect\citeauthoryear{{Capanema}, {Esmaili}  \&
  {Serpico}}{{Capanema} et~al.}{2021}]{Capanema2021}
{Capanema} A.,  {Esmaili} A.,   {Serpico} P.~D.,  2021, \mn@doi [\jcap]
  {10.1088/1475-7516/2021/02/037}, \href
  {https://ui.adsabs.harvard.edu/abs/2021JCAP...02..037C} {2021, 037}

\bibitem[\protect\citeauthoryear{{Caprioli}, {Haggerty}  \& {Blasi}}{{Caprioli}
  et~al.}{2020}]{Caprioli2020}
{Caprioli} D.,  {Haggerty} C.~C.,   {Blasi} P.,  2020, \mn@doi [\apj]
  {10.3847/1538-4357/abbe05}, \href
  {https://ui.adsabs.harvard.edu/abs/2020ApJ...905....2C} {905, 2}

\bibitem[\protect\citeauthoryear{{Carilli}}{{Carilli}}{1996}]{Carilli_2006}
{Carilli} C.~L.,  1996, Astronomy and Astrophysics, \href
  {http://adsabs.harvard.edu/abs/1996A%26A...305..402C} {305, 402}

\bibitem[\protect\citeauthoryear{{Castor}, {McCray}  \& {Weaver}}{{Castor}
  et~al.}{1975}]{Castor-1976}
{Castor} J.,  {McCray} R.,   {Weaver} R.,  1975, \mn@doi [\apjl]
  {10.1086/181908}, \href
  {https://ui.adsabs.harvard.edu/abs/1975ApJ...200L.107C} {200, L107}

\bibitem[\protect\citeauthoryear{{Chevalier} \& {Clegg}}{{Chevalier} \&
  {Clegg}}{1985}]{ChevalierClegg85}
{Chevalier} R.~A.,  {Clegg} A.~W.,  1985, \mn@doi [\nat] {10.1038/317044a0},
  \href {https://ui.adsabs.harvard.edu/abs/1985Natur.317...44C} {317, 44}

\bibitem[\protect\citeauthoryear{{Cicone} et~al.,}{{Cicone}
  et~al.}{2014}]{Cicone-2014}
{Cicone} C.,  et~al., 2014, \mn@doi [\aap] {10.1051/0004-6361/201322464}, \href
  {https://ui.adsabs.harvard.edu/abs/2014A&A...562A..21C} {562, A21}

\bibitem[\protect\citeauthoryear{{Cooper}, {Bicknell}, {Sutherland}  \&
  {Bland-Hawthorn}}{{Cooper} et~al.}{2007}]{Cooper2007}
{Cooper} J.~L.,  {Bicknell} G.~V.,  {Sutherland} R.~S.,   {Bland-Hawthorn} J.,
  2007, \mn@doi [\apss] {10.1007/s10509-007-9526-4}, \href
  {https://ui.adsabs.harvard.edu/abs/2007Ap&SS.311...99C} {311, 99}

\bibitem[\protect\citeauthoryear{{Di Matteo}, {Bournaud}, {Martig}, {Combes},
  {Melchior}  \& {Semelin}}{{Di Matteo} et~al.}{2008}]{DiMatteo2008}
{Di Matteo} P.,  {Bournaud} F.,  {Martig} M.,  {Combes} F.,  {Melchior} A.~L.,
   {Semelin} B.,  2008, \mn@doi [Astronomy \& Astrophysics]
  {10.1051/0004-6361:200809480}, \href
  {https://ui.adsabs.harvard.edu/abs/2008A&A...492...31D} {492, 31}

\bibitem[\protect\citeauthoryear{{Dorfi} \& {Breitschwerdt}}{{Dorfi} \&
  {Breitschwerdt}}{2012}]{Dorfi2012}
{Dorfi} E.~A.,  {Breitschwerdt} D.,  2012, \mn@doi [\aap]
  {10.1051/0004-6361/201118082}, \href
  {https://ui.adsabs.harvard.edu/abs/2012A&A...540A..77D} {540, A77}

\bibitem[\protect\citeauthoryear{{Dundovic}, {Pezzi}, {Blasi}, {Evoli}  \&
  {Matthaeus}}{{Dundovic} et~al.}{2020}]{Dundovic2020}
{Dundovic} A.,  {Pezzi} O.,  {Blasi} P.,  {Evoli} C.,   {Matthaeus} W.~H.,
  2020, \mn@doi [\prd] {10.1103/PhysRevD.102.103016}, \href
  {https://ui.adsabs.harvard.edu/abs/2020PhRvD.102j3016D} {102, 103016}

\bibitem[\protect\citeauthoryear{{Epimakhov} et~al.}{{Epimakhov}
  et~al.}{2013}]{Tunka2013}
{Epimakhov} S.,  et~al., 2013, in International Cosmic Ray Conference. p.~818

\bibitem[\protect\citeauthoryear{{Everett}, {Zweibel}, {Benjamin}, {McCammon},
  {Rocks}  \& {Gallagher}}{{Everett} et~al.}{2008}]{Everett2008}
{Everett} J.~E.,  {Zweibel} E.~G.,  {Benjamin} R.~A.,  {McCammon} D.,  {Rocks}
  L.,   {Gallagher} John~S. I.,  2008, \mn@doi [\apj] {10.1086/524766}, \href
  {https://ui.adsabs.harvard.edu/abs/2008ApJ...674..258E} {674, 258}

\bibitem[\protect\citeauthoryear{{Fielding}, {Quataert}, {Martizzi}  \&
  {Faucher-Gigu{\`e}re}}{{Fielding} et~al.}{2017}]{Fielding2017}
{Fielding} D.,  {Quataert} E.,  {Martizzi} D.,   {Faucher-Gigu{\`e}re} C.-A.,
  2017, \mn@doi [\mnras] {10.1093/mnrasl/slx072}, \href
  {https://ui.adsabs.harvard.edu/abs/2017MNRAS.470L..39F} {470, L39}

\bibitem[\protect\citeauthoryear{{Fielding}, {Quataert}  \&
  {Martizzi}}{{Fielding} et~al.}{2018}]{Fielding2018}
{Fielding} D.,  {Quataert} E.,   {Martizzi} D.,  2018, \mn@doi [\mnras]
  {10.1093/mnras/sty2466}, \href
  {https://ui.adsabs.harvard.edu/abs/2018MNRAS.481.3325F} {481, 3325}

\bibitem[\protect\citeauthoryear{{F{\"o}rster Schreiber} et~al.}{{F{\"o}rster
  Schreiber} et~al.}{2001}]{ForsterSchreiber_2001}
{F{\"o}rster Schreiber} N.~M.,  et~al., 2001, \mn@doi [The Astrophysical
  Journal] {10.1086/320546}, \href
  {http://adsabs.harvard.edu/abs/2001ApJ...552..544F} {552, 544}

\bibitem[\protect\citeauthoryear{{Franceschini} \& {Rodighiero}}{{Franceschini}
  \& {Rodighiero}}{2017}]{Franceschini-EBL}
{Franceschini} A.,  {Rodighiero} G.,  2017, \mn@doi [\aap]
  {10.1051/0004-6361/201629684}, \href
  {https://ui.adsabs.harvard.edu/abs/2017A&A...603A..34F} {603, A34}

\bibitem[\protect\citeauthoryear{{Galliano}, {Dwek}  \& {Chanial}}{{Galliano}
  et~al.}{2008}]{Galliano2008}
{Galliano} F.,  {Dwek} E.,   {Chanial} P.,  2008, \mn@doi [\apj]
  {10.1086/523621}, \href
  {https://ui.adsabs.harvard.edu/abs/2008ApJ...672..214G} {672, 214}

\bibitem[\protect\citeauthoryear{{Gao} \& {Solomon}}{{Gao} \&
  {Solomon}}{2004}]{Gao_Solomon_2004}
{Gao} Y.,  {Solomon} P.~M.,  2004, \mn@doi [The Astrophysical Journal]
  {10.1086/382999}, \href {http://adsabs.harvard.edu/abs/2004ApJ...606..271G}
  {606, 271}

\bibitem[\protect\citeauthoryear{{Girichidis}, {Pfrommer}, {Pakmor}  \&
  {Springel}}{{Girichidis} et~al.}{2021}]{Girichidis2021}
{Girichidis} P.,  {Pfrommer} C.,  {Pakmor} R.,   {Springel} V.,  2021, arXiv
  e-prints, \href {https://ui.adsabs.harvard.edu/abs/2021arXiv210913250G} {p.
  arXiv:2109.13250}

\bibitem[\protect\citeauthoryear{{Gruppioni} et~al.,}{{Gruppioni}
  et~al.}{2015}]{Gruppioni_2015}
{Gruppioni} C.,  et~al., 2015, \mn@doi [\mnras] {10.1093/mnras/stv1204}, \href
  {https://ui.adsabs.harvard.edu/abs/2015MNRAS.451.3419G} {451, 3419}

\bibitem[\protect\citeauthoryear{Haack \& Wiebusch}{Haack \&
  Wiebusch}{2018}]{IceCube_muons}
Haack C.,  Wiebusch C.,  2018, \mn@doi [PoS] {10.22323/1.301.1005}, ICRC2017,
  1005

\bibitem[\protect\citeauthoryear{{Hanasz}, {Lesch}, {Naab}, {Gawryszczak},
  {Kowalik}  \& {W{\'o}lta{\'n}ski}}{{Hanasz} et~al.}{2013}]{Hanasz2013}
{Hanasz} M.,  {Lesch} H.,  {Naab} T.,  {Gawryszczak} A.,  {Kowalik} K.,
  {W{\'o}lta{\'n}ski} D.,  2013, \mn@doi [\apjl] {10.1088/2041-8205/777/2/L38},
  \href {https://ui.adsabs.harvard.edu/abs/2013ApJ...777L..38H} {777, L38}

\bibitem[\protect\citeauthoryear{IceCube Collaboration:~Aartsen et~al.}{IceCube
  Collaboration:~Aartsen et~al.}{2013}]{First_Ice_nu}
IceCube Collaboration:~Aartsen M.,  et~al., 2013, \mn@doi [Science]
  {10.1126/science.1242856}, 342

\bibitem[\protect\citeauthoryear{{Kachelrie{\ss}} \&
  {Semikoz}}{{Kachelrie{\ss}} \& {Semikoz}}{2006}]{Kach}
{Kachelrie{\ss}} M.,  {Semikoz} D.~V.,  2006, \mn@doi [Physics Letters B]
  {10.1016/j.physletb.2006.01.009}, \href
  {https://ui.adsabs.harvard.edu/abs/2006PhLB..634..143K} {634, 143}

\bibitem[\protect\citeauthoryear{{Kafexhiu}, {Aharonian}, {Taylor}  \&
  {Vila}}{{Kafexhiu} et~al.}{2014}]{kafexhiu2014}
{Kafexhiu} E.,  {Aharonian} F.,  {Taylor} A.~M.,   {Vila} G.~S.,  2014, \mn@doi
  [\prd] {10.1103/PhysRevD.90.123014}, \href
  {https://ui.adsabs.harvard.edu/abs/2014PhRvD..90l3014K} {90, 123014}

\bibitem[\protect\citeauthoryear{{Kelner} \& {Aharonian}}{{Kelner} \&
  {Aharonian}}{2008}]{Kelner-photomeson}
{Kelner} S.~R.,  {Aharonian} F.~A.,  2008, \mn@doi [\prd]
  {10.1103/PhysRevD.78.034013}, \href
  {https://ui.adsabs.harvard.edu/abs/2008PhRvD..78c4013K} {78, 034013}

\bibitem[\protect\citeauthoryear{Kelner, Aharonian  \& Bugayov}{Kelner
  et~al.}{2006}]{Kelner_Aharonian_2006_proton-proton}
Kelner S.~R.,  Aharonian F.~A.,   Bugayov V.~V.,  2006, \mn@doi [Phys. Rev.]
  {10.1103/PhysRevD.74.034018, 10.1103/PhysRevD.79.039901}, D74, 034018

\bibitem[\protect\citeauthoryear{Kennicutt}{Kennicutt}{1998}]{Kennicutt:1998zb}
Kennicutt Jr. R.~C.,  1998, \mn@doi [Ann. Rev. Astron. Astrophys.]
  {10.1146/annurev.astro.36.1.189}, 36, 189

\bibitem[\protect\citeauthoryear{{Koo} \& {McKee}}{{Koo} \&
  {McKee}}{1992a}]{Koo-McKee}
{Koo} B.-C.,  {McKee} C.~F.,  1992a, \mn@doi [The Astrophysical Journal]
  {10.1086/171132}, \href
  {https://ui.adsabs.harvard.edu/abs/1992ApJ...388...93K} {388, 93}

\bibitem[\protect\citeauthoryear{{Koo} \& {McKee}}{{Koo} \&
  {McKee}}{1992b}]{Koo-McKee2}
{Koo} B.-C.,  {McKee} C.~F.,  1992b, \mn@doi [\apj] {10.1086/171133}, \href
  {https://ui.adsabs.harvard.edu/abs/1992ApJ...388..103K} {388, 103}

\bibitem[\protect\citeauthoryear{{Kornecki}, {Pellizza}, {del Palacio},
  {M{\"u}ller}, {Albacete-Colombo}  \& {Romero}}{{Kornecki}
  et~al.}{2020}]{Kornecki2020}
{Kornecki} P.,  {Pellizza} L.~J.,  {del Palacio} S.,  {M{\"u}ller} A.~L.,
  {Albacete-Colombo} J.~F.,   {Romero} G.~E.,  2020, \mn@doi [\aap]
  {10.1051/0004-6361/202038428}, \href
  {https://ui.adsabs.harvard.edu/abs/2020A&A...641A.147K} {641, A147}

\bibitem[\protect\citeauthoryear{{Kornecki}, {Peretti}, {del Palacio},
  {Benaglia}  \& {Pellizza}}{{Kornecki} et~al.}{2021}]{Kornecki2021}
{Kornecki} P.,  {Peretti} E.,  {del Palacio} S.,  {Benaglia} P.,   {Pellizza}
  L.~J.,  2021, arXiv e-prints, \href
  {https://ui.adsabs.harvard.edu/abs/2021arXiv210700823K} {p. arXiv:2107.00823}

\bibitem[\protect\citeauthoryear{{Krumholz}, {Crocker}, {Xu}, {Lazarian},
  {Rosevear}  \& {Bedwell-Wilson}}{{Krumholz} et~al.}{2020}]{Krumholz2020}
{Krumholz} M.~R.,  {Crocker} R.~M.,  {Xu} S.,  {Lazarian} A.,  {Rosevear}
  M.~T.,   {Bedwell-Wilson} J.,  2020, \mn@doi [\mnras]
  {10.1093/mnras/staa493}, \href
  {https://ui.adsabs.harvard.edu/abs/2020MNRAS.493.2817K} {493, 2817}

\bibitem[\protect\citeauthoryear{{Lamastra} et~al.,}{{Lamastra}
  et~al.}{2016}]{Lamastra1}
{Lamastra} A.,  et~al., 2016, \mn@doi [\aap] {10.1051/0004-6361/201628667},
  \href {https://ui.adsabs.harvard.edu/abs/2016A&A...596A..68L} {596, A68}

\bibitem[\protect\citeauthoryear{{Lamastra}, {Tavecchio}, {Romano}, {Landoni}
  \& {Vercellone}}{{Lamastra} et~al.}{2019}]{Lamastra2}
{Lamastra} A.,  {Tavecchio} F.,  {Romano} P.,  {Landoni} M.,   {Vercellone} S.,
   2019, \mn@doi [Astroparticle Physics] {10.1016/j.astropartphys.2019.04.003},
  \href {https://ui.adsabs.harvard.edu/abs/2019APh...112...16L} {112, 16}

\bibitem[\protect\citeauthoryear{{Linden}}{{Linden}}{2017}]{Linden_2017}
{Linden} T.,  2017, \mn@doi [\prd] {10.1103/PhysRevD.96.083001}, \href
  {https://ui.adsabs.harvard.edu/abs/2017PhRvD..96h3001L} {96, 083001}

\bibitem[\protect\citeauthoryear{Lisanti, Mishra-Sharma, Necib  \&
  Safdi}{Lisanti et~al.}{2016}]{Lisanti_2016}
Lisanti M.,  Mishra-Sharma S.,  Necib L.,   Safdi B.~R.,  2016, \mn@doi [The
  Astrophysical Journal] {10.3847/0004-637x/832/2/117}, 832, 117

\bibitem[\protect\citeauthoryear{{Liu}, {Murase}, {Inoue}, {Ge}  \&
  {Wang}}{{Liu} et~al.}{2018}]{Liu2018}
{Liu} R.-Y.,  {Murase} K.,  {Inoue} S.,  {Ge} C.,   {Wang} X.-Y.,  2018,
  \mn@doi [\apj] {10.3847/1538-4357/aaba74}, \href
  {https://ui.adsabs.harvard.edu/abs/2018ApJ...858....9L} {858, 9}

\bibitem[\protect\citeauthoryear{{Lochhaas}, {Thompson}, {Quataert}  \&
  {Weinberg}}{{Lochhaas} et~al.}{2018}]{Lochaas2018}
{Lochhaas} C.,  {Thompson} T.~A.,  {Quataert} E.,   {Weinberg} D.~H.,  2018,
  \mn@doi [\mnras] {10.1093/mnras/sty2421}, \href
  {https://ui.adsabs.harvard.edu/abs/2018MNRAS.481.1873L} {481, 1873}

\bibitem[\protect\citeauthoryear{{Loeb} \& {Waxman}}{{Loeb} \&
  {Waxman}}{2006}]{Loeb-Waxmann2006}
{Loeb} A.,  {Waxman} E.,  2006, \mn@doi [\jcap]
  {10.1088/1475-7516/2006/05/003}, \href
  {https://ui.adsabs.harvard.edu/abs/2006JCAP...05..003L} {2006, 003}

\bibitem[\protect\citeauthoryear{{Mannucci} et~al.}{{Mannucci}
  et~al.}{2003}]{Mannucci_etal_2003}
{Mannucci} F.,  et~al., 2003, \mn@doi [Astronomy and Astrophysics]
  {10.1051/0004-6361:20030198}, \href
  {http://adsabs.harvard.edu/abs/2003A\%26A...401..519M} {401, 519}

\bibitem[\protect\citeauthoryear{{McQuinn}, {Skillman}, {Cannon}, {Dalcanton},
  {Dolphin}, {Stark}  \& {Weisz}}{{McQuinn} et~al.}{2009}]{McQuinn2009}
{McQuinn} K. B.~W.,  {Skillman} E.~D.,  {Cannon} J.~M.,  {Dalcanton} J.~J.,
  {Dolphin} A.,  {Stark} D.,   {Weisz} D.,  2009, \mn@doi [The Astrophysical
  Journal] {10.1088/0004-637X/695/1/561}, \href
  {https://ui.adsabs.harvard.edu/abs/2009ApJ...695..561M} {695, 561}

\bibitem[\protect\citeauthoryear{{Melioli}, {de Gouveia Dal Pino}  \&
  {Geraissate}}{{Melioli} et~al.}{2013}]{Elizabete_2013}
{Melioli} C.,  {de Gouveia Dal Pino} E.~M.,   {Geraissate} F.~G.,  2013,
  \mn@doi [\mnras] {10.1093/mnras/stt126}, \href
  {https://ui.adsabs.harvard.edu/abs/2013MNRAS.430.3235M} {430, 3235}

\bibitem[\protect\citeauthoryear{{Merten}, {Bustard}, {Zweibel}  \& {Becker
  Tjus}}{{Merten} et~al.}{2018}]{Merten_2018}
{Merten} L.,  {Bustard} C.,  {Zweibel} E.~G.,   {Becker Tjus} J.,  2018,
  \mn@doi [\apj] {10.3847/1538-4357/aabfdd}, \href
  {https://ui.adsabs.harvard.edu/abs/2018ApJ...859...63M} {859, 63}

\bibitem[\protect\citeauthoryear{{Morlino}, {Blasi}, {Peretti}  \&
  {Cristofari}}{{Morlino} et~al.}{2021}]{MBPC2021}
{Morlino} G.,  {Blasi} P.,  {Peretti} E.,   {Cristofari} P.,  2021, arXiv
  e-prints, \href {https://ui.adsabs.harvard.edu/abs/2021arXiv210209217M} {p.
  arXiv:2102.09217}

\bibitem[\protect\citeauthoryear{{M{\"u}ller}, {Romero}  \&
  {Roth}}{{M{\"u}ller} et~al.}{2020}]{Muller-Romero}
{M{\"u}ller} A.~L.,  {Romero} G.~E.,   {Roth} M.,  2020, \mn@doi [MNRAS]
  {10.1093/mnras/staa1720}, \href
  {https://ui.adsabs.harvard.edu/abs/2020MNRAS.496.2474M} {496, 2474}

\bibitem[\protect\citeauthoryear{{Murase}, {Guetta}  \& {Ahlers}}{{Murase}
  et~al.}{2016}]{Murase-hidden-CR-ACC}
{Murase} K.,  {Guetta} D.,   {Ahlers} M.,  2016, \mn@doi [\prl]
  {10.1103/PhysRevLett.116.071101}, \href
  {https://ui.adsabs.harvard.edu/abs/2016PhRvL.116g1101M} {116, 071101}

\bibitem[\protect\citeauthoryear{{Murase}, {Kimura}  \&
  {M{\'e}sz{\'a}ros}}{{Murase} et~al.}{2020}]{Murase:2020}
{Murase} K.,  {Kimura} S.~S.,   {M{\'e}sz{\'a}ros} P.,  2020, \mn@doi [\prl]
  {10.1103/PhysRevLett.125.011101}, \href
  {https://ui.adsabs.harvard.edu/abs/2020PhRvL.125a1101M} {125, 011101}

\bibitem[\protect\citeauthoryear{Palladino, Fedynitch, Rasmussen  \&
  Taylor}{Palladino et~al.}{2018}]{Palladino-starburst:2018}
Palladino A.,  Fedynitch A.,  Rasmussen R.~W.,   Taylor A.~M.,  2018, arXiv
  e-prints, p. arXiv:1812.04685

\bibitem[\protect\citeauthoryear{{Papadopoulos}, {Thi}, {Miniati}  \&
  {Viti}}{{Papadopoulos} et~al.}{2011}]{Papadopoulos}
{Papadopoulos} P.~P.,  {Thi} W.-F.,  {Miniati} F.,   {Viti} S.,  2011, \mn@doi
  [\mnras] {10.1111/j.1365-2966.2011.18504.x}, \href
  {https://ui.adsabs.harvard.edu/abs/2011MNRAS.414.1705P} {414, 1705}

\bibitem[\protect\citeauthoryear{{Park} \& {VERITAS Collaboration}}{{Park} \&
  {VERITAS Collaboration}}{2015}]{Veritas-sensitivity}
{Park} N.,  {VERITAS Collaboration} 2015, in 34th International Cosmic Ray
  Conference (ICRC2015). p.~771 (\mn@eprint {arXiv} {1508.07070})

\bibitem[\protect\citeauthoryear{Peng et~al.}{Peng et~al.}{2016}]{Peng:2016nsx}
Peng F.-K.,  et~al., 2016, \mn@doi [The Astrophysical Journal]
  {10.3847/2041-8205/821/2/L20}, 821, L20

\bibitem[\protect\citeauthoryear{{Peretti}, {Blasi}, {Aharonian}  \&
  {Morlino}}{{Peretti} et~al.}{2019}]{Peretti-1}
{Peretti} E.,  {Blasi} P.,  {Aharonian} F.,   {Morlino} G.,  2019, \mn@doi
  [\mnras] {10.1093/mnras/stz1161}, \href
  {https://ui.adsabs.harvard.edu/abs/2019MNRAS.487..168P} {487, 168}

\bibitem[\protect\citeauthoryear{{Peretti}, {Blasi}, {Aharonian}, {Morlino}  \&
  {Cristofari}}{{Peretti} et~al.}{2020}]{Peretti-2}
{Peretti} E.,  {Blasi} P.,  {Aharonian} F.,  {Morlino} G.,   {Cristofari} P.,
  2020, \mn@doi [\mnras] {10.1093/mnras/staa698}, \href
  {https://ui.adsabs.harvard.edu/abs/2020MNRAS.493.5880P} {493, 5880}

\bibitem[\protect\citeauthoryear{{Pfrommer}, {Pakmor}, {Simpson}  \&
  {Springel}}{{Pfrommer} et~al.}{2017}]{Pfrommer2017}
{Pfrommer} C.,  {Pakmor} R.,  {Simpson} C.~M.,   {Springel} V.,  2017, \mn@doi
  [\apjl] {10.3847/2041-8213/aa8bb1}, \href
  {https://ui.adsabs.harvard.edu/abs/2017ApJ...847L..13P} {847, L13}

\bibitem[\protect\citeauthoryear{{Prosin} et~al.}{{Prosin}
  et~al.}{2016}]{Tunka2016}
{Prosin} V.~V.,  et~al., 2016, in European Physical Journal Web of Conferences.
  p. 03004, \mn@doi{10.1051/epjconf/201612103004}

\bibitem[\protect\citeauthoryear{{Recchia}, {Blasi}  \& {Morlino}}{{Recchia}
  et~al.}{2016}]{Recchia2016}
{Recchia} S.,  {Blasi} P.,   {Morlino} G.,  2016, \mn@doi [\mnras]
  {10.1093/mnras/stw1966}, \href
  {https://ui.adsabs.harvard.edu/abs/2016MNRAS.462.4227R} {462, 4227}

\bibitem[\protect\citeauthoryear{{Recchia}, {Gabici}, {Aharonian}  \&
  {Niro}}{{Recchia} et~al.}{2021}]{Recchia-M31}
{Recchia} S.,  {Gabici} S.,  {Aharonian} F.~A.,   {Niro} V.,  2021, arXiv
  e-prints, \href {https://ui.adsabs.harvard.edu/abs/2021arXiv210105016R} {p.
  arXiv:2101.05016}

\bibitem[\protect\citeauthoryear{{Romero}, {M{\"u}ller}  \& {Roth}}{{Romero}
  et~al.}{2018}]{Romero-Muller}
{Romero} G.~E.,  {M{\"u}ller} A.~L.,   {Roth} M.,  2018, \mn@doi [Astronomy \&
  Astrophysics] {10.1051/0004-6361/201832666}, \href
  {https://ui.adsabs.harvard.edu/abs/2018A&A...616A..57R} {616, A57}

\bibitem[\protect\citeauthoryear{{Roth}, {Krumholz}, {Crocker}  \&
  {Celli}}{{Roth} et~al.}{2021}]{Roth-nature}
{Roth} M.~A.,  {Krumholz} M.~R.,  {Crocker} R.~M.,   {Celli} S.,  2021, \mn@doi
  [\nat] {10.1038/s41586-021-03802-x}, \href
  {https://ui.adsabs.harvard.edu/abs/2021Natur.597..341R} {597, 341}

\bibitem[\protect\citeauthoryear{{Schneider}, {Ostriker}, {Robertson}  \&
  {Thompson}}{{Schneider} et~al.}{2020}]{Schneider2020}
{Schneider} E.~E.,  {Ostriker} E.~C.,  {Robertson} B.~E.,   {Thompson} T.~A.,
  2020, \mn@doi [\apj] {10.3847/1538-4357/ab8ae8}, \href
  {https://ui.adsabs.harvard.edu/abs/2020ApJ...895...43S} {895, 43}

\bibitem[\protect\citeauthoryear{{Strickland} \& {Heckman}}{{Strickland} \&
  {Heckman}}{2009}]{Strickland-Heckmann2009}
{Strickland} D.~K.,  {Heckman} T.~M.,  2009, \mn@doi [\apj]
  {10.1088/0004-637X/697/2/2030}, \href
  {https://ui.adsabs.harvard.edu/abs/2009ApJ...697.2030S} {697, 2030}

\bibitem[\protect\citeauthoryear{{Strickland} \& {Stevens}}{{Strickland} \&
  {Stevens}}{2000}]{Strickland-Stevens-2000}
{Strickland} D.~K.,  {Stevens} I.~R.,  2000, \mn@doi [\mnras]
  {10.1046/j.1365-8711.2000.03391.x}, \href
  {https://ui.adsabs.harvard.edu/abs/2000MNRAS.314..511S} {314, 511}

\bibitem[\protect\citeauthoryear{{Strickland}, {Heckman}, {Weaver}, {Hoopes}
  \& {Dahlem}}{{Strickland} et~al.}{2002}]{Strickland_2002}
{Strickland} D.~K.,  {Heckman} T.~M.,  {Weaver} K.~A.,  {Hoopes} C.~G.,
  {Dahlem} M.,  2002, \mn@doi [\apj] {10.1086/338889}, \href
  {https://ui.adsabs.harvard.edu/abs/2002ApJ...568..689S} {568, 689}

\bibitem[\protect\citeauthoryear{{Subedi} et~al.,}{{Subedi}
  et~al.}{2017}]{Subedi2017}
{Subedi} P.,  et~al., 2017, \mn@doi [\apj] {10.3847/1538-4357/aa603a}, \href
  {https://ui.adsabs.harvard.edu/abs/2017ApJ...837..140S} {837, 140}

\bibitem[\protect\citeauthoryear{{Sudoh}, {Totani}  \& {Kawanaka}}{{Sudoh}
  et~al.}{2018}]{Sudoh}
{Sudoh} T.,  {Totani} T.,   {Kawanaka} N.,  2018, \mn@doi [\pasj]
  {10.1093/pasj/psy039}, \href
  {https://ui.adsabs.harvard.edu/abs/2018PASJ...70...49S} {70, 49}

\bibitem[\protect\citeauthoryear{Tamborra, Ando  \& Murase}{Tamborra
  et~al.}{2014}]{Tamborra-Ando-Murase:2014}
Tamborra I.,  Ando S.,   Murase K.,  2014, \mn@doi [JCAP]
  {10.1088/1475-7516/2014/09/043}, 1409, 043

\bibitem[\protect\citeauthoryear{{Taylor}, {Gabici}  \& {Aharonian}}{{Taylor}
  et~al.}{2014}]{Taylor-2014}
{Taylor} A.~M.,  {Gabici} S.,   {Aharonian} F.,  2014, \mn@doi [\prd]
  {10.1103/PhysRevD.89.103003}, \href
  {https://ui.adsabs.harvard.edu/abs/2014PhRvD..89j3003T} {89, 103003}

\bibitem[\protect\citeauthoryear{Tenorio-Tagle \&
  {Mu\~noz-Tu\~n\'on}}{Tenorio-Tagle \&
  {Mu\~noz-Tu\~n\'on}}{1997}]{Tenorio_Tagle_1997}
Tenorio-Tagle G.,  {Mu\~noz-Tu\~n\'on} C.,  1997, \mn@doi [The Astrophysical
  Journal] {10.1086/303780}, 478, 134

\bibitem[\protect\citeauthoryear{{Thompson} et~al.}{{Thompson}
  et~al.}{2006}]{Thompson_etal_2006}
{Thompson} T.~A.,  et~al., 2006, \mn@doi [The Astrophysical Journal]
  {10.1086/504035}, \href {http://adsabs.harvard.edu/abs/2006ApJ...645..186T}
  {645, 186}

\bibitem[\protect\citeauthoryear{{Tumlinson}, {Peeples}  \& {Werk}}{{Tumlinson}
  et~al.}{2017}]{Tumlinson2017}
{Tumlinson} J.,  {Peeples} M.~S.,   {Werk} J.~K.,  2017, \mn@doi [\araa]
  {10.1146/annurev-astro-091916-055240}, \href
  {https://ui.adsabs.harvard.edu/abs/2017ARA&A..55..389T} {55, 389}

\bibitem[\protect\citeauthoryear{{Veilleux}, {Cecil}  \&
  {Bland-Hawthorn}}{{Veilleux} et~al.}{2005}]{Veilleux}
{Veilleux} S.,  {Cecil} G.,   {Bland-Hawthorn} J.,  2005, \mn@doi [Annual
  Review of Astronomy \& Astrophysics]
  {10.1146/annurev.astro.43.072103.150610}, \href
  {https://ui.adsabs.harvard.edu/abs/2005ARA&A..43..769V} {43, 769}

\bibitem[\protect\citeauthoryear{{Vercellone}}{{Vercellone}}{2016}]{Astri}
{Vercellone} S.,  2016, in European Physical Journal Web of Conferences. p.
  04006 (\mn@eprint {arXiv} {1508.00799}),
  \mn@doi{10.1051/epjconf/201612104006}

\bibitem[\protect\citeauthoryear{{Wang} \& {Loeb}}{{Wang} \&
  {Loeb}}{2017}]{Wang-Loeb2017}
{Wang} X.,  {Loeb} A.,  2017, \mn@doi [\prd] {10.1103/PhysRevD.95.063007},
  \href {https://ui.adsabs.harvard.edu/abs/2017PhRvD..95f3007W} {95, 063007}

\bibitem[\protect\citeauthoryear{{Weaver}, {McCray}, {Castor}, {Shapiro}  \&
  {Moore}}{{Weaver} et~al.}{1977}]{Weaver77}
{Weaver} R.,  {McCray} R.,  {Castor} J.,  {Shapiro} P.,   {Moore} R.,  1977,
  \mn@doi [The Astrophysical Journal] {10.1086/155692}, \href
  {https://ui.adsabs.harvard.edu/abs/1977ApJ...218..377W} {218, 377}

\bibitem[\protect\citeauthoryear{{Werhahn}, {Pfrommer}, {Girichidis}  \&
  {Winner}}{{Werhahn} et~al.}{2021}]{Werhahn2021}
{Werhahn} M.,  {Pfrommer} C.,  {Girichidis} P.,   {Winner} G.,  2021, \mn@doi
  [\mnras] {10.1093/mnras/stab1325}, \href
  {https://ui.adsabs.harvard.edu/abs/2021MNRAS.505.3295W} {505, 3295}

\bibitem[\protect\citeauthoryear{{Westmoquette}, {Smith}, {Gallagher},
  {Trancho}, {Bastian}  \& {Konstantopoulos}}{{Westmoquette}
  et~al.}{2009a}]{Westmoquette1}
{Westmoquette} M.~S.,  {Smith} L.~J.,  {Gallagher} J.~S. I.,  {Trancho} G.,
  {Bastian} N.,   {Konstantopoulos} I.~S.,  2009a, \mn@doi [\apj]
  {10.1088/0004-637X/696/1/192}, \href
  {https://ui.adsabs.harvard.edu/abs/2009ApJ...696..192W} {696, 192}

\bibitem[\protect\citeauthoryear{{Westmoquette}, {Gallagher}, {Smith},
  {Trancho}, {Bastian}  \& {Konstantopoulos}}{{Westmoquette}
  et~al.}{2009b}]{Westmoquette2}
{Westmoquette} M.~S.,  {Gallagher} J.~S.,  {Smith} L.~J.,  {Trancho} G.,
  {Bastian} N.,   {Konstantopoulos} I.~S.,  2009b, \mn@doi [\apj]
  {10.1088/0004-637X/706/2/1571}, \href
  {https://ui.adsabs.harvard.edu/abs/2009ApJ...706.1571W} {706, 1571}

\bibitem[\protect\citeauthoryear{{Wik} et~al.}{{Wik}
  et~al.}{2014}]{Wik_NGC253_2014}
{Wik} D.~R.,  et~al., 2014, \mn@doi [The Astrophysical Journal]
  {10.1088/0004-637X/797/2/79}, \href
  {http://adsabs.harvard.edu/abs/2014ApJ...797...79W} {797, 79}

\bibitem[\protect\citeauthoryear{{Williams} \& {Bower}}{{Williams} \&
  {Bower}}{2010}]{Radio_Williams_Bower_2010}
{Williams} P.~K.~G.,  {Bower} G.~C.,  2010, \mn@doi [The Astrophysical Journal]
  {10.1088/0004-637X/710/2/1462}, \href
  {http://adsabs.harvard.edu/abs/2010ApJ...710.1462W} {710, 1462}

\bibitem[\protect\citeauthoryear{Yoast-Hull, Everett, Gallagher  \&
  Zweibel}{Yoast-Hull et~al.}{2013}]{Yoast-Hull_M82_2013}
Yoast-Hull T.~M.,  Everett J.~E.,  Gallagher J.~S.,   Zweibel E.~G.,  2013,
  \mn@doi [Astrophys. J.] {10.1088/0004-637X/768/1/53}, 768, 53

\bibitem[\protect\citeauthoryear{{Zhang}}{{Zhang}}{2018}]{Zhang2018}
{Zhang} D.,  2018, \mn@doi [Galaxies] {10.3390/galaxies6040114}, \href
  {https://ui.adsabs.harvard.edu/abs/2018Galax...6..114Z} {6, 114}

\bibitem[\protect\citeauthoryear{{Zhang}, {Murase}  \&
  {M{\'e}sz{\'a}ros}}{{Zhang} et~al.}{2020}]{Zhang+Murase2020}
{Zhang} Z.,  {Murase} K.,   {M{\'e}sz{\'a}ros} P.,  2020, \mn@doi [\mnras]
  {10.1093/mnras/staa022}, \href
  {https://ui.adsabs.harvard.edu/abs/2020MNRAS.492.2250Z} {492, 2250}

\bibitem[\protect\citeauthoryear{Zirakashvili \& Völk}{Zirakashvili \&
  Völk}{2006}]{ZIRAKASHVILI20061923}
Zirakashvili V.,  Völk H.,  2006, \mn@doi [Advances in Space Research]
  {https://doi.org/10.1016/j.asr.2005.06.013}, 37, 1923

\bibitem[\protect\citeauthoryear{{Zorn}}{{Zorn}}{2019}]{HESS}
{Zorn} J.,  2019, in 36th International Cosmic Ray Conference (ICRC2019).
  p.~834 (\mn@eprint {arXiv} {1908.04620})

\makeatother
\end{thebibliography}




\appendix

\section{Estimates of time-scales and luminosity of the shocked bubble}
\label{Appendix: Bubble-lumin}

The order of magnitude of the wind bubble luminosity can be estimated in a simplified way from the total power of the system and the dominant timescales.
The dominant escape time from the system is the advection which reads
\begin{equation}
    \tau_{\rm adv} \approx \frac{R_{\rm esc}}{\langle v_2 \rangle} \approx 10^2 \, R_{\rm FS,1} u_{1,3}^{-1} \, \rm Myr
\end{equation}
where $\langle v_2 \rangle \sim u_2/3$ is the average wind speed in the downstream region and $R_{\rm FS,1}$ is the forward shock location in units of 10 kpc. 

The loss timescales for pp and $p\gamma$ collisions are described by the following expressions
\begin{eqnarray}
    \tau_{\rm pp} = [\xi_{\rm pp} \, n_2 \, \sigma_{\rm pp} \, c ]^{-1} \approx 4 \cdot 10^3 \, n_{2,-2}^{-1} \, \rm Myr \\
    \tau_{\rm p\gamma} = [\xi_{p\gamma} \, n_{ph} \, \sigma_{\rm p\gamma} \, c ]^{-1} \approx 10^3 \, U_{\rm OPT,3}^{-1} \, \chi_{1}^{-2} \, \rm Myr 
\end{eqnarray}
where we assumed that the target photon field is a $\delta$--Dirac at 1 eV and where $\chi= R_{\rm SBN}/R_{\rm sh}$, while $\xi_{\rm pp} \sim 0.1$ and $\xi_{\rm p \gamma} \sim 0.2$ are the elasticity factors. 

Since $\tau_{\rm adv} \ll \tau_{\rm loss}$ one concludes that the dynamical relevance of losses is negligible so that, the assumption of negligible energy losses is fully justified in the shocked wind. 

Since the advection dominates, the CR distribution function is approximately constant in the whole shocked bubble.
Assuming now that the luminosity of accelerated particles is a fraction $\xi_{\rm CR}$ of the total wind power $\Dot{E}= \, \Dot{M} \, u_1^2/2$, we can express the gamma--ray luminosity as
\begin{equation}
    L_{\gamma} \approx \frac{\xi_{\rm CR}}{2} \Dot{M}\,u_1^2 \, \frac{R_{\rm esc}}{\langle v_2 \rangle} \, n_2 \sigma_{\rm pp} c \xi_{\rm pp}
\end{equation}
which converted in units of standard parameters can be rewritten as
\begin{equation}
    L_{\gamma} \approx 2 \cdot 10^{40} \,  \Dot{M}_1 \, u_{1,8} \, R_{\rm FS,1} \, n_{2,-2} \, \xi_{\rm CR,-1} \, \rm erg \, s^{-1}.
\end{equation}
Interestingly, such a value is a fraction $\gtrsim 10 \%$ of the luminosity of an SBN when a supernova rate of $0.1 \, \rm yr^{-1} $ and perfect calorimetric conditions are assumed. Indeed, under these conditions, the nucleus luminosity in gamma--rays can be written as:
\begin{equation}
    L_{\rm SBN} \approx \mathcal{R}_{\rm SN} \xi_{\rm CR} E_{\rm SN} \approx 1.6 \cdot 10^{41} \mathcal{R}_{\rm SN,-1} \, \xi_{\rm CR,-1} \, E_{\rm SN,51}.
\end{equation}
However, differently from the SBN, the gamma--ray luminosity of the bubble increases with time.

\section{Energy loss functions for accelerated particles in spherical geometry}
\label{Appendix: Calculations}

We report here the analytic expressions of the functions included in the solution presented in \S~\ref{Sec3: Model}.

The function $G_1$, embedding the adiabatic energy loss/gain has the following expression:
\begin{equation}
    \label{Eq: G-function}
    G_1(r,p)= \frac{1}{3} \int_{0}^{r} dr' \, \Tilde{q}(r',p) \, f_1(r',p) \, \partial_{r'}[r'^2 \, u_1(r')]
\end{equation}
where $\Tilde{q}$ reads
\begin{equation}
    \label{Eq: q_tilde-function}
    \Tilde{q}(r,p)= -\frac{\partial{\rm ln}p^3f(r,p)}{\partial {\rm ln}p }.
\end{equation}

The function $H_1$ accounts for pp energy losses and reads
\begin{equation}
    \label{Eq: H-function}
    H_1(r,p)= \int_{0}^{r} dr' \, r'^2 \, f_1(r',p) \, n(r') \, \sigma_{\rm pp}(p) \, c \, ,
\end{equation}
where $\sigma_{\rm pp}$ is the pp cross section \citep[see][]{Kelner_Aharonian_2006_proton-proton}.


\bsp	
\label{lastpage}
\end{document}